\documentclass[preprint,aps ,nofootinbib]{revtex4}
\usepackage{graphicx}
\usepackage{amsmath}
\usepackage{amsfonts}
\usepackage{amssymb}
\usepackage{color}%
\usepackage{dcolumn}
\providecommand{\U}[1]{\protect\rule{.1in}{.1in}}

\newcommand{\f}{\begin{equation}}
\newcommand{\ff}{\end{equation}}
\newcommand{\fa}{\begin{eqnarray}}
\newcommand{\ffa}{\end{eqnarray}}

\begin{document}
\title{ Holographic Superconductors in Quasi-topological Gravity}
\author{Xiao-Mei Kuang}
\email{xmeikuang@gmail.com}
\author{Wei-Jia Li}
\email{li831415@163.com}
\author{Yi Ling}
\email{yling@ncu.edu.cn} \affiliation{ Center for Relativistic
Astrophysics and High Energy Physics, Department of Physics,
Nanchang University, 330031, China}

\begin{abstract}
In this paper we study (3+1) dimensional holographic
superconductors in quasi-topological gravity which is recently
proposed by R. Myers {\it et.al.}. Through both analytical and
numerical analysis, we find in general the condensation becomes
harder with the increase of coupling parameters of higher
curvature terms. In particular, comparing with those in ordinary
Gauss-Bonnet gravity, we find that positive cubic corrections in
quasi-topological gravity suppress the condensation while negative
cubic terms make it easier. We also calculate the conductivity
numerically for various coupling parameters. It turns out that the
universal relation of $\omega_g/T_c\simeq 8$ is unstable and this
ratio becomes larger with the increase of the coupling parameters.
A brief discussion on the condensation from the CFT side is also
presented.
\end{abstract} \maketitle
\section {Introduction}
The AdS/CFT correspondence links a ($d+1$) dimensional
 gravity theory in the bulk to a dual $d$ dimensional
quantum field theory on the
boundary\cite{ADS,WittenADS,MaldacenaReviewADS}. Recent progress
indicates that this duality plays a significant role in studying
various phenomena in condensed matter
physics\cite{Hartnoll1,Hartnoll2,Minic,Hartnoll3,Hartnoll4,McGreevy},
 since it provides a powerful tool to describe some
 strongly correlated behaviors in the vicinity of quantum critical point
 which has not been well understood in
conventional field theory. Particularly, in \cite{Gubser1,Gubser2}
Gubser shows the possibility that there is a spontaneous breaking
of U(1) symmetry near the AdS black hole horizon in the bulk when
the Higgs model couples to gravity. This mechanism leads to a
phenomenon that AdS black holes can be surrounded by a
nonvanishing scalar field $\psi$ when the temperature is low
enough, which calls for a re-understanding of the `no hair'
theorem of black hole. More importantly, based on gauge/gravity
correspondence such a process of forming scalar hairs strongly
implies a second-order phase transition occurring on the CFT side.
Inspired by this observation, Hartnoll \textit{et al.} constructed
a holographic superconductor model exhibiting such a phase
transition\cite{Hartnoll5}. It is remarkable that in this model
some basic features of superconductors can be accomplished below a
critical temperature. Stimulated by this work, various holographic
superconductors have been constructed in other gravity theories,
for instance the Gauss-Bonnet (GB) gravity, M theory and
Ho\v{r}ava-Lifshitz theory
etc.\cite{Albash,Gubser3,Wen,Gauntlett,Gregory,Sin,Cai,Ge,Pan,Wu,Cai2}.

Based on AdS/CFT correspondence, the presence of higher order
derivatives in AdS gravity means new couplings among operators in
the dual CFT. So various higher curvature couplings will lead to
more classes of dual field theories. However, in ordinary GB
gravity only one quadratic coupling term involves, which greatly
limits the range of dual field theories. In order to extend
holographic studies to new classes of CFTs, we may consider
introducing new terms with higher order derivatives into gravity,
for instance a curvature-cubed term. A straightforward way is to
add the cubic term in Lovelock gravity. Unfortunately, such a term
is topological and has effects on equations of motion (EOM) only
in very high dimensions. Until recently, Myers \textit{et al.}
construct a new higher derivative theory of gravity in 5
dimensional spacetime which contains not only the Gauss-Bonnet
term but also a curvature-cubed
interaction\cite{Myers}.\footnote{It is worthwhile to point out
that before their work some important insights into the
curvature-cubed interaction has been given in \cite{oliva}. Recent
relevant work can also been found in
\cite{ray,sinha1,sinha2,amsel}.} Unlike Lovelock gravity, this
cubic term is not purely topological but dynamically contributive
to the evolution of fields in the bulk, thus it may be useful for
us to apply the holographic studies to wider classes of field
theories in ordinary dimension. In gauge/gravity duality,
quasi-topological gravity theory is thought to be dual to the
large $N$ limit of some conformal field theory without
supersymmetry. In this theory equations of motion are the fourth
order in derivatives in general backgrounds. However, the
linearized equations reduce to the second order in
$\textrm{AdS}_5$ case, which implies the holographic description
in this model is still under control. Moreover, exact black hole
solutions with an asymptotically AdS behavior have been found and
many interesting features have been revealed when different
coupling parameters are chosen\cite{Myers}. The holographic
studies for these black hole solutions have been started and some
recipes of AdS/CFT dictionary for this duality have been presented
in \cite{Myers2}. Applying this to the hydrodynamics of dual
fields, they found that the curvature-cubed term suppresses the
ratio bound of the shear viscosity to entropy density to
$\eta/s\approx\frac{0.4140}{4\pi}$, which is conjectured to be
$\frac{1}{4\pi} $ in Einstein's gravity. In this paper, we intend
to continue the holographic studies on quasi-topological gravity
by constructing a holographic model of superconductors in the
probe limit. Employing both analytical and numerical methods, we
obtain the critical temperature $T_c$ of the superconductors, and
find that in general the condensation becomes harder with the
increase of coupling parameters of higher curvature terms. In
particular, comparing with those in ordinary Gauss-Bonnet gravity,
we find that positive cubic corrections in quasi-topological
gravity suppress the occurrence of phase transitions while
negative cubic terms make it easier. Furthermore, we calculate the
conductivity $\sigma$ numerically and find the universal relation
of $\omega_g/T_c\simeq 8$ is unstable and the ratio becomes larger
with the increase of coupling parameters. According to the gauge/gravity duality,
we find the dual description with direct parameters in the field theory on the boundary.

The paper is organized as follows. In section II, we quickly review
the black hole physics in quasi-topological gravity. Then focusing
on the black hole solution with positive coupling parameters we
construct a holographic superconductor model and study the
condensation of the scalar field analytically and numerically. In
section III, the conductivity is calculated numerically and the
ratio of the frequency gap to the critical temperature is discussed.
We turn to investigate the holographic superconductivity for black
hole solutions with negative coupling parameters in section IV. Its
dual picture on the CFT side is also briefly discussed. Our
conclusions and discussions are presented in the last section.

\section {Quasi-topological Holographic Superconductors}

In quasi-topological gravity, the bulk action of gravity in
five-dimensional spacetime is given by \cite{Myers}
\begin{equation}\label{a}
S_g=\frac{1}{16\pi G_5}\int
d^5x\sqrt{-g}\Big[R+\frac{12}{L^2}+\frac{\alpha
L^2}{2}\mathcal{X}_4+\frac{7\beta L^4}{8}\mathcal {Z}_5\Big],
\end{equation}
where $\alpha$ and $\beta$ are Gauss-Bonnet coupling parameter and
curvature-cubed interaction parameter, respectively. Here
$\mathcal{X}_4$ is defined as
\begin{equation}\label{a1}
\mathcal{X}_4=R_{\mu\nu\rho\sigma}R^{\mu\nu\rho\sigma}-4R_{\mu\nu}R^{\mu\nu}+R^2,
\end{equation}
and $\mathcal{Z}_5$ is a curvature-cubed term with the form
\begin{eqnarray}\label{a2}
\mathcal{Z}_5={R_{\mu\nu}}^{\rho\sigma}{R_{\rho\sigma}}^{\alpha\beta}{R_{\alpha
\beta}}^{\mu\nu}+\frac{1}{14}(21R_{\mu\nu\rho\sigma}{R^{\mu\nu\rho\sigma}}R
-120R_{\mu\nu\rho\sigma}{R^{\mu\nu\rho}}_{\alpha}R^{\sigma\alpha}\nonumber\\
+144R_{\mu\nu\rho\sigma}R^{\mu\rho}R^{\nu\sigma}
+128{R_\mu}^{\nu}{R_\nu}^{\rho}{R_\rho}^{\mu}-108{R_\mu}^{\nu}{R_\nu}^{\mu}R+11R^3).
\end{eqnarray}
The planar black hole solution with an asymptotically AdS behavior
in quasi-topological gravity has been obtained in \cite{Myers} as
well. It is given as
\begin{equation}\label{b}
ds^2=\frac{r^2}{L^2}(-N(r)^2f(r)dt^2+dx^2+dy^2+dz^2)+\frac{L^2}{r^2f(r)}dr^2,
\end{equation}
where $N(r)$ is the lapse function. Inserting this metric into
(\ref{a}) and taking the variation $\delta N$, one finds that
$f(r)$ should satisfy the following equation
\begin{equation}\label{c}
1-f(r)+\alpha f(r)^2+\beta f(r)^3=\frac{\eta^4}{r^4},
\end{equation}
where $\eta$ is a constant and $f(r)$ vanishes on the horizon of
the black hole $r_H$, namely $f(r_H)=0$. Following
Ref.\cite{Myers}, we choose $\eta=r_H$ and $N^2=1/f_\infty$, where
$f_\infty$ satisfies
\begin{equation}\label{c1}
1-f_\infty+\alpha f_\infty^2+\beta f_\infty^3=0.
\end{equation}
Usually a black hole horizon exists if $f(r\neq r_H)>0$, so with
positive $\alpha$ and $\beta$ we have
\begin{equation}\label{c2}
f_\infty=1+\alpha f_\infty^2+\beta f_\infty^3>1,
\end{equation}
which means the lapse function $N$ is always smaller than one in
this case. With the standard approach, one can find the Hawking
temperature of the black hole is given by
\begin{equation}\label{d}
T=\frac{N}{4\pi}f'(r)|_{r=r_H}=\frac{Nr_H}{\pi L^2}.
\end{equation}
It will also be viewed as the temperature of the dual CFT on the
boundary.

In the bulk of this black hole background, the action of matter is
proposed as
\begin{equation}\label{e}
S_m=\int d^5x\sqrt{-g}\Big[-\frac{1}{4}F^{\mu \nu}F_{\mu
\nu}-|\nabla \psi-iqA \psi|^2-m^2|\psi|^2\Big],
\end{equation}
where $F_{\mu\nu}$ is the strength of a gauge field with U(1)
symmetry and $\psi$ is a charged scalar field. Through this paper
we only consider the probe limit, which means the backreaction of
matter fields on the background can be neglected if the charge is
large enough. Taking the radial ansatz, $\psi=\psi(r)$ and
$A_{\mu}=(\phi(r),0,0,0,0)$ and rescaling two fields by a factor
$q^{-2}$, we obtain the EOM of $\psi$ and $\phi$ as
\begin{equation}\label{f}
\psi''+(\frac{g'}{g}+\frac{3}{r})\psi'+(\frac{\phi^2}{N^2g^2}-\frac{m^2}{L^2g})\psi=0
\end{equation}
and
\begin{equation}\label{g}
\phi''+\frac{3}{r}\phi'-\frac{2\psi^2}{g}\phi=0,
\end{equation}
where $g(r)=\frac{r^2f(r)}{L^2}$ and the prime denotes a
derivative with respect to $r$. In the following analysis, we
choose the mass term $m^{2}=-3/L^2$ which is above the
Breitenlohner-Freedman bound for initial
stability\cite{Breitenlohner}. In order to solve these equations
we need to give some boundary conditions at the horizon and the
asymptotically AdS region $(r\rightarrow\infty)$:

\noindent $\bullet$ The regularity conditions at the horizon
($r=r_H$) give rise to
\begin{eqnarray}\label{h}
\phi(r_H)=0,\hspace{1cm} \psi(r_H)=\frac{g'\mid
_{r=r_H}}{m^2}r_H\psi^\prime(r_H) \ .
\end{eqnarray}
\noindent $\bullet$ Near the AdS boundary ($r\rightarrow\infty$),
the asymptotical behaviors of fields are like
\begin{eqnarray}\label{i}
\phi(r)=\mu - \frac{\rho}{r^2}\,,\hspace{0.5cm}
\psi=\frac{C_{-}}{r^{\lambda_-}}+\frac{C_{+}}{r^{\lambda_+}}\,,
\label{r:boundary}
\end{eqnarray}
where $\lambda_\pm=2\pm\sqrt{4-3\frac{\tilde{L}^2}{L^2}}$ with
 an effective AdS radius $\tilde{L}=L/\sqrt{f_\infty}=L N$. In the above equation $\mu$ and $\rho$
are understood as the chemical potential and the charge density on
the boundary respectively. From Eq.(\ref{c2}), it is easy to find
that the value of $\lambda_+$ in quasi-topological gravity is
larger than that in standard Einstein and Gauss-Bonnet theory.

Next, we need to find non-trivial solutions to the scalar field
$\psi$ in the bulk. Such kind of black hole hair is possible since
gravity has a negative contribution to the effective mass term which
leads to a breaking of the U(1) gauge symmetry. As the scalar field
changes from $\psi=0$ to $\psi\neq0$ in the bulk, correspondingly
one thinks that a phase transition occurs around a critical
temperature $T_c$ for the dual conformal field theory on the
boundary.  More explicitly, according to the recipe of AdS/CFT
dictionary, the non-zero scalar field breaking the local U(1)
symmetry in the bulk is actually dual to a condensation operator
(order parameter) $\mathcal {O}$ which breaks the global U(1)
symmetry in the large N field on the boundary, and the crucial
relations are $\langle\mathcal {O_+}\rangle\sim C_+$ and
$\langle\mathcal {O_-}\rangle\sim C_-$, respectively. Here $C_-$ and
$C_+$ can also be considered as the source and the vacuum
expectation values of the operator. Therefore, a non-vanishing
$\psi$ implies the occurrence of the condensation in CFT below the
critical temperature. Next we set $C_-=0$ and investigate the
condensation of $\langle\mathcal {O_+}\rangle$ from analytical and
numerical aspect separately. We remark that the other case of
setting $C_+=0$ can be worked out in a similar manner.
\section*{A. Analytical calculation}
At first we will
 analytically calculate the critical temperature for the
phase transition of the holographic superconductors, and then
demonstrate how the critical temperature behaves as coupling
parameters $\alpha$ and $\beta$ change.

In this section we focus our discussion on positive coupling
parameters. As pointed out in Ref.\cite{Myers}, if both coupling
parameters $\alpha$ and $\beta$ are positive, only one real root
of (\ref{c}) leads to a stable AdS black hole solution. It has the
form as
\begin{equation}\label{j}
f_3(r)=-\frac{1}{2}(u+v)-i\frac{\sqrt{3}}{2}(u-v)-\frac{\alpha}{3\beta},
\end{equation}
where
\begin{eqnarray}\label{k}
u=(q+\sqrt{q^2-p^3})^{1/3}
,\hspace{1cm}v=(q-\sqrt{q^2-p^3})^{1/3},\label{n}\end{eqnarray}
and
\begin{eqnarray}
p=\frac{3\beta+\alpha^2}{9\beta^2},\hspace{1cm}q=-\frac{2\alpha^3+9\alpha\beta+27\beta^2(1-\frac{r_H^4}{r^4})}{54\beta^3}.
\end{eqnarray}
\begin{figure}
\center{
\includegraphics[scale=0.6]{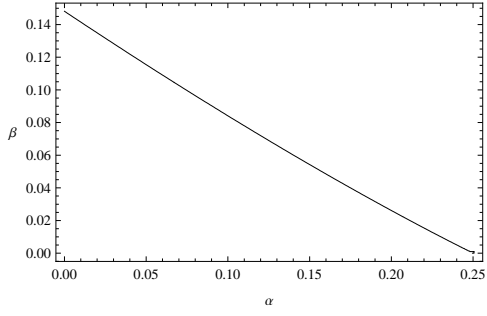}\\ \hspace{2cm}
\caption{Below the border line is the range of $\alpha$ and
$\beta$ which maintains the black hole stable.}}
\end{figure}

To maintain this AdS solution stable, one more condition should be
 satisfied, which is
$\mathcal {D}_\infty=(q^2-p^3)|_{r\rightarrow\infty}<0$. This
condition confines the values of $\alpha$ and $\beta$ to be
chosen. In FIG. 1 we illustrate a border line for $\alpha$ vs.
$\beta$, only below which this black hole solution makes sense.
From this figure, we also notice that the larger the value of
$\alpha$ is fixed, the smaller the range of $\beta$ could take,
and vice versa. Especially, if we let $\beta=0$, the largest value
of $\alpha$ would be $0.25$, which is nothing but the Chern-Simons
limit in GB gravity. Moreover, there also exists a limit
$\beta=\frac{4}{27}$ when we take $\alpha=0$. Keep this in mind,
we will make all the following analysis in an appropriate range
which is below the upper bound of $\alpha$ and $\beta$.

For simplicity, we define a dimensionless variable
$z=\frac{r_H}{r}$. Then the EOM of two fields can be rewritten as
\begin{equation}\label{l}
\psi''+(\frac{g'}{g}-\frac{1}{z})\psi'+\frac{r_{H}^2}{z^4}\Big(\frac{\phi^2}{N^2g^2}+\frac{3}{L^2g}\Big)=0
\end{equation}
and
\begin{equation}\label{m}
\phi''-\frac{1}{z}\phi'-\frac{r_H^2}{z^4}\frac{2\psi^2}{g}\phi=0,
\end{equation}
where the prime denotes a derivative with respect to $z$ now. The
corresponding boundary conditions can be expressed as follows.

\noindent $\bullet$ At the horizon ($z=1$)
\begin{equation}\label{n}
\phi(1)=0,\hspace{1cm}\psi(1)=\frac{g'\mid _{z=1}}{3}\psi^\prime(1)
\ .
\end{equation}
\noindent $\bullet$ Near the boundary ($z\rightarrow0$)
\begin{equation}\label{o}
\phi=\mu-\rho
\frac{z^2}{r_H^2}\equiv\mu-qz^2,\hspace{1cm}\psi=C_+z^{\lambda_+}+C_-z^{\lambda_-},
\end{equation}
We take the Taylor expansion of the fields near the horizon as:
\begin{equation}\label{p}
\phi(z)=\phi(1)-\phi'(1)(1-z)+\frac{1}{2}\phi''(1)(1-z)^2+\cdot\cdot\cdot
,
\end{equation}
\begin{equation}\label{q}
\psi(z)=\psi(1)-\psi'(1)(1-z)+\frac{1}{2}\psi''(1)(1-z)^2+\cdot\cdot\cdot.
\end{equation}
Then expanding (\ref{f}) and (\ref{g}) near the horizon gives
\begin{equation}\label{r}
\phi''(1)\approx \Big(1-\frac{L^2}{2}\psi(1)^2\Big)\phi'(1) ,
\end{equation}
\begin{equation}\label{r1}
\psi''(1)\approx
\Big(-\frac{15}{8}\psi'(1)+4\alpha\Big)\psi'(1)-\frac{L^4}{32N^2r_H^2}\phi'(1)^2\psi(1).
\end{equation}
We obtain the following important expressions by inserting
(\ref{r}), (\ref{r1}) into (\ref{p}) and (\ref{q}), respectively
\begin{eqnarray}
&&\phi(z)\approx-\phi'(1)(1-z)+\frac{1}{2}\Big(1-\frac{L^2}{2}\psi(1)^2\Big)\phi'(1)(1-z)^2+\cdot\cdot\cdot
,\label{s}\\
&&\psi(z)\approx
\psi(1)-\frac{3}{4}\psi(1)(1-z)+\Big(-\frac{15}{64}+\frac{3\alpha}{2}-\frac{L^4}{64N^2r_H^2}\phi'(1)^2\Big)\psi(1)(1-z)^2+\cdot\cdot\cdot.\label{s1}
\end{eqnarray}
Now, matching the solutions (\ref{o}), (\ref{s}) and (\ref{s1}) at
any point $z_m$ in the region $z\in[0,1]$, for instance
$z_m=\frac{1}{2}$, we obtain four continuity equations of wave
functions
\begin{eqnarray}\label{t1}
&&\mu-\frac{1}{4}q=\frac{1}{2}b-\frac{1}{8}b\left(1-\frac{L^2}{2}a^2\right)
\\\label{t2}
&&-q=-b+\frac{1}{2}b\left(1-\frac{L^2}{2}a^2\right)
\\\label{t3}
&&C_+\left(\frac{1}{2}\right)^{\lambda_+}=\frac{5}{8}a+\frac{1}{4}a
\left(-\frac{15}{64}+\frac{3\alpha}{2}-\frac{L^4}{64N^2r_H^2}b^2\right)
\\\label{t4}
&&2\lambda_+C_+\left(\frac{1}{2}\right)^{\lambda_+}=\frac{3}{4}a
-a\left(-\frac{15}{64}+\frac{3\alpha}{2}-\frac{L^4}{64N^2r_H^2}b^2
\right),
\end{eqnarray}
where we have defined $\psi(1)=a$ and $-\phi'(1)=b$ for simplicity.
Combining (\ref{t3}) and (\ref{t4}), we obtain
\begin{equation}\label{u}
C_+=\frac{13}{8}\frac{2^{\lambda_+}}{\lambda_++2}a,
\end{equation}
where $a$ can be solved from (\ref{t2})
\begin{equation}\label{v}
a^2=\frac{4q}{L^2b}\left(1-\frac{b}{2q}\right).
\end{equation}
Moreover, combining (\ref{t3}) and (\ref{u}) we get
\begin{equation}\label{w}
b=8\frac{Nr_H}{L^2} \sqrt{\frac{5\lambda_+-3}
{2(\lambda_++2)}-\frac{15}{64}+\frac{3\alpha}{2}}.
\end{equation}
Then (\ref{v}) can be rewritten as
\begin{equation}\label{x}
a^2=\frac{2}{L^2}\frac{T_c^3}{T^3}\left(1-\frac{T^3}{T_c^3}\right),
\end{equation}
where the critical temperature $T_c$ is defined as
\begin{equation}\label{y}
T_c=\Big[\frac{\rho}{4L}\frac{1}{\sqrt{\frac{5\lambda_+-3}
{2(\lambda_++2)}-\frac{15}{64}+\frac{3\alpha}{2}}}\Big]^{\frac{1}{3}}\frac{N^{\frac{2}{3}}}{\pi
L}.
\end{equation}
Since $N$ is smaller than one and
$\frac{5\lambda_+-3}{2(\lambda_++2)}$ is a monotonic
 increasing function of $\lambda_+$, we find the
value of $T_c$ depends on couplings and decreases as the value of
$\alpha$ or $\beta$ increases. Thus we may conclude that the
presence of the positive curvature-cubed term makes the
condensation more difficult. Moreover, based on AdS/CFT dictionary
$\langle {\cal O_+} \rangle\equiv L C_{+} r_H^{\lambda_+}
L^{-2\lambda_{+}}$, we express the condensation as
\begin{equation}\label{z}
\frac{\langle {\cal O_+} \rangle^{\frac{1}{\lambda_+}}}{T_c}
=\frac{2\pi}{N} \left(\frac{13}{8}\frac{\sqrt{2}}{\lambda_++2}
\right)^{\frac{1}{\lambda_+}}\frac{T}{T_c}
\left[~\frac{T_c^3}{T^3}\left(1-\frac{T^3}{T_c^3}\right)
~\right]^{\frac{1}{2\lambda_+}}.
\end{equation}
Therefore the condensation $\langle\mathcal {O_+}\rangle\neq 0$
only if $T<T_c$, and its critical behavior $(1-T^3/T_c^3)^{1/2}$
implies the phase transition is the second order. To present a
more qualitative analysis about the impact of the curvature-cubed
term on the condensation, we compare critical temperatures of
condensation with those in standard Einstein gravity and $GB$
gravity by substituting specific values of $\alpha$ and $\beta$
into (\ref{y}).

In Einstein's gravity, one has $T_c=0.2017\rho^{1/3}/L$ by
analytical calculation. In GB gravity if one sets $\alpha=0.01$,
the critical temperature
$T_c=0.2004\rho^{1/3}/L$.\footnote{Different from
Ref.\cite{Gregory}, we have set the lapse function
$N=1/\sqrt{f_\infty}$ rather than one, so that in our case the
induced (3+1) metric on the boundary is conformal to the Minkowski
space. Therefore the value of $T_c$ obtained here is a little bit
smaller than that obtained in \cite{Gregory}.} While in
quasi-topological gravity, once the Gauss-Bonnet constant $\alpha$
is fixed, the critical temperature goes down as the
curvature-cubed constant $\beta$ increases. For instance if we fix
$\alpha=0.01$, but increase $\beta$ from 0.001 to 0.01 and 0.1,
$T_c$ decreases from $T_c=0.2003\rho^{1/3}$ to $0.1995\rho^{1/3}$
and $0.1889\rho^{1/3}$. Similarly, fixing $\beta=0.001$ but
changing $\alpha$ from 0.0001 to 0.1 and 0.2, one has $T_c$ from
$0.2016\rho^{1/3}$ to $0.1880\rho^{1/3}$ and $0.1653\rho^{1/3}$.
The explicit dependence of $T_c$ on coupling parameters $\alpha$
and $\beta$ is shown in FIG.2 and FIG.3. In FIG.3, we notice that
$T_c$ experiences a quick-fall at the right-up corner when
$\alpha$ and $\beta$ are large enough, implying that the
analytical result (\ref{y}) is no longer stable in this region,
which is consistent with our previous discussion about the range
of $\alpha$ and $\beta$ as shown in FIG.1.

\begin{figure}
\center{
\includegraphics[scale=0.65]{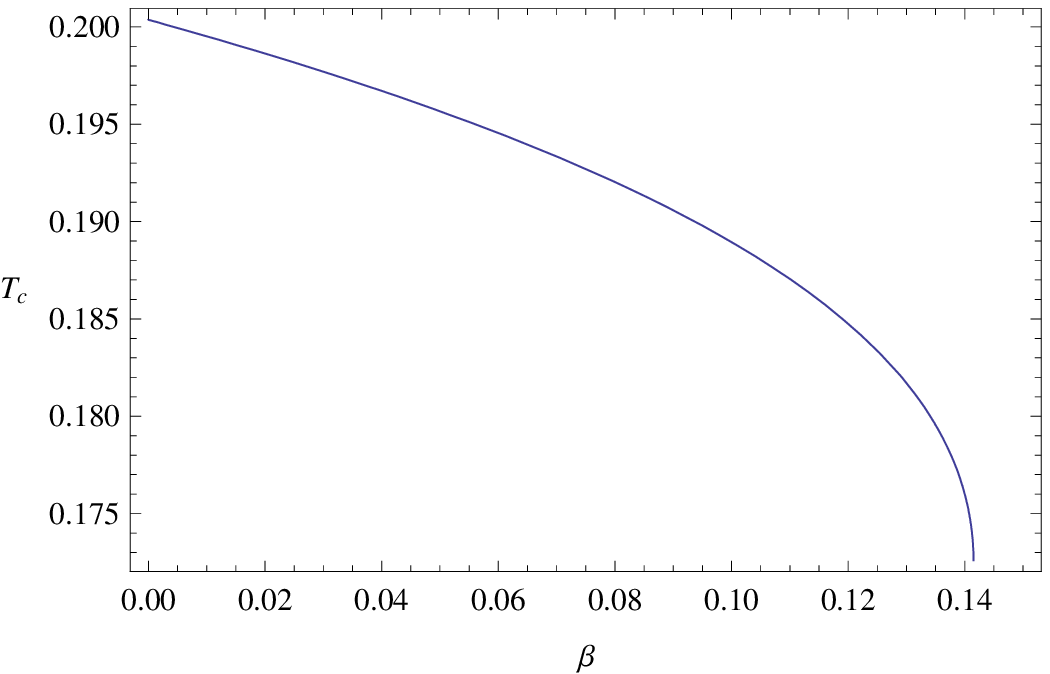}\hspace{1cm}
\includegraphics[scale=0.65]{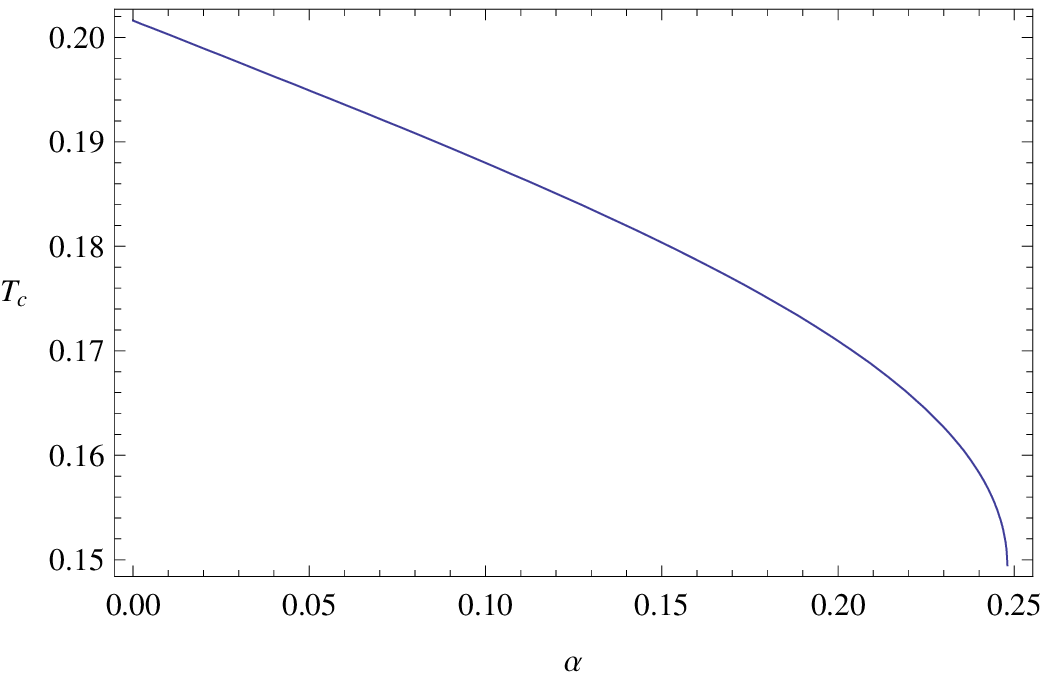}\\ \hspace{1cm}
\caption{ Relations between $T_c$ and the coupling parameters
$\alpha$ and $\beta$, respectively. In the left figure the line is
for $\alpha=0.01$, while in the right figure the line is for
$\beta=0.001$.}}
\end{figure}
\begin{figure}
\center{
\includegraphics[scale=0.6]{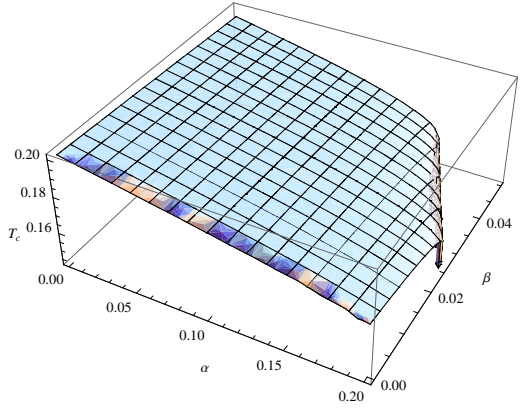}\\ \hspace{2cm}
\caption{The critical temperature $T_c$ as a function of $\alpha$
and $\beta$.}}
\end{figure}

At the end of this subsection we should mention that the
analytical approximation given above works well only when the
temperature is not too far from $T_c$. In order to have a complete
understanding on the relation between the condensation and
temperature we need to solve equations of motion in a numerical
way. That is what we intend to do in the next subsection.

\section*{B. Numerical result}
We solve equations of motion (\ref{f}) and (\ref{g}) numerically and
plot FIG.4 to demonstrate the condensation as a function of
temperature.
\begin{figure} \center{
\includegraphics[scale=0.4]{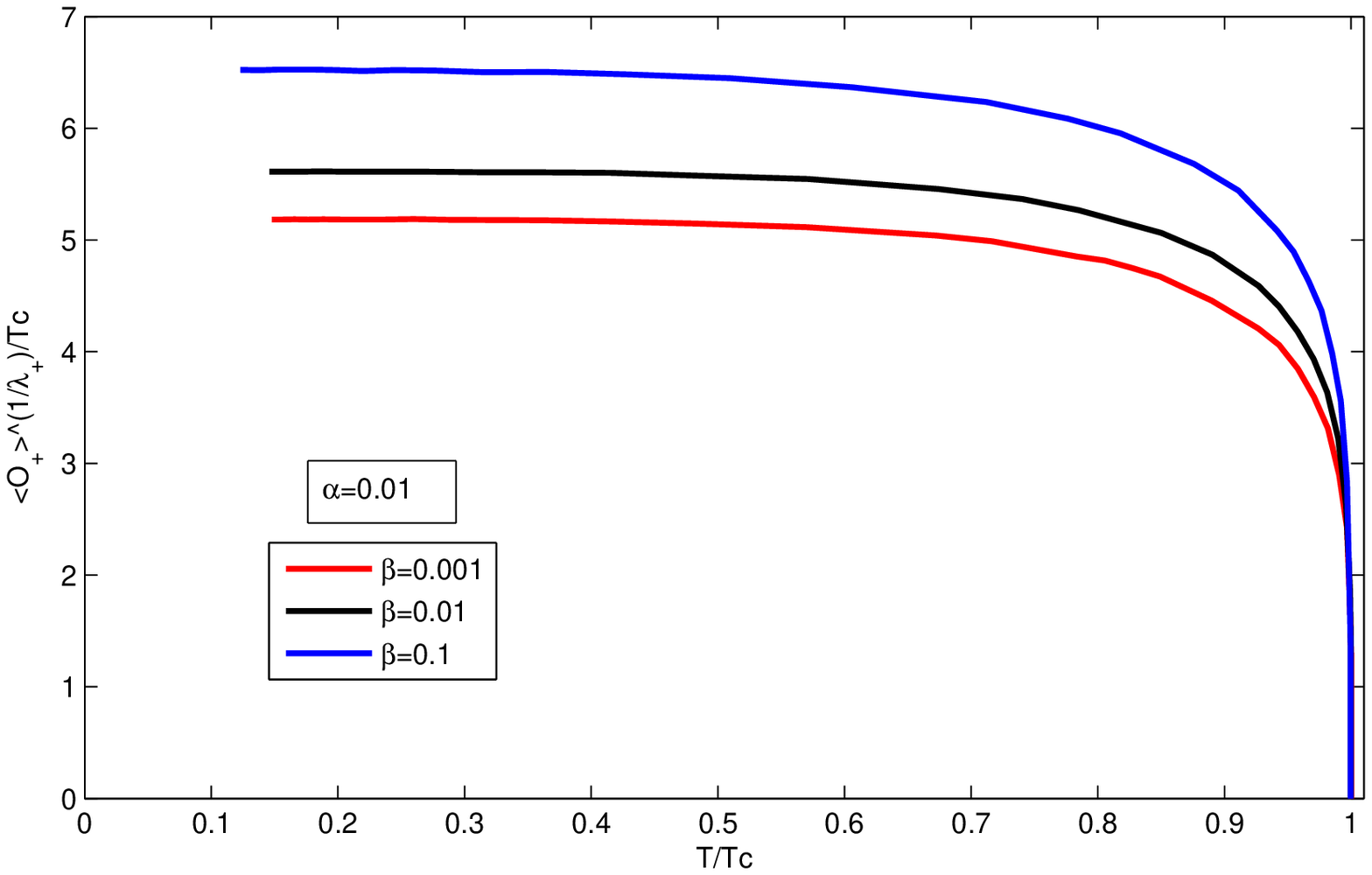}\hspace{0cm}
\includegraphics[scale=0.37]{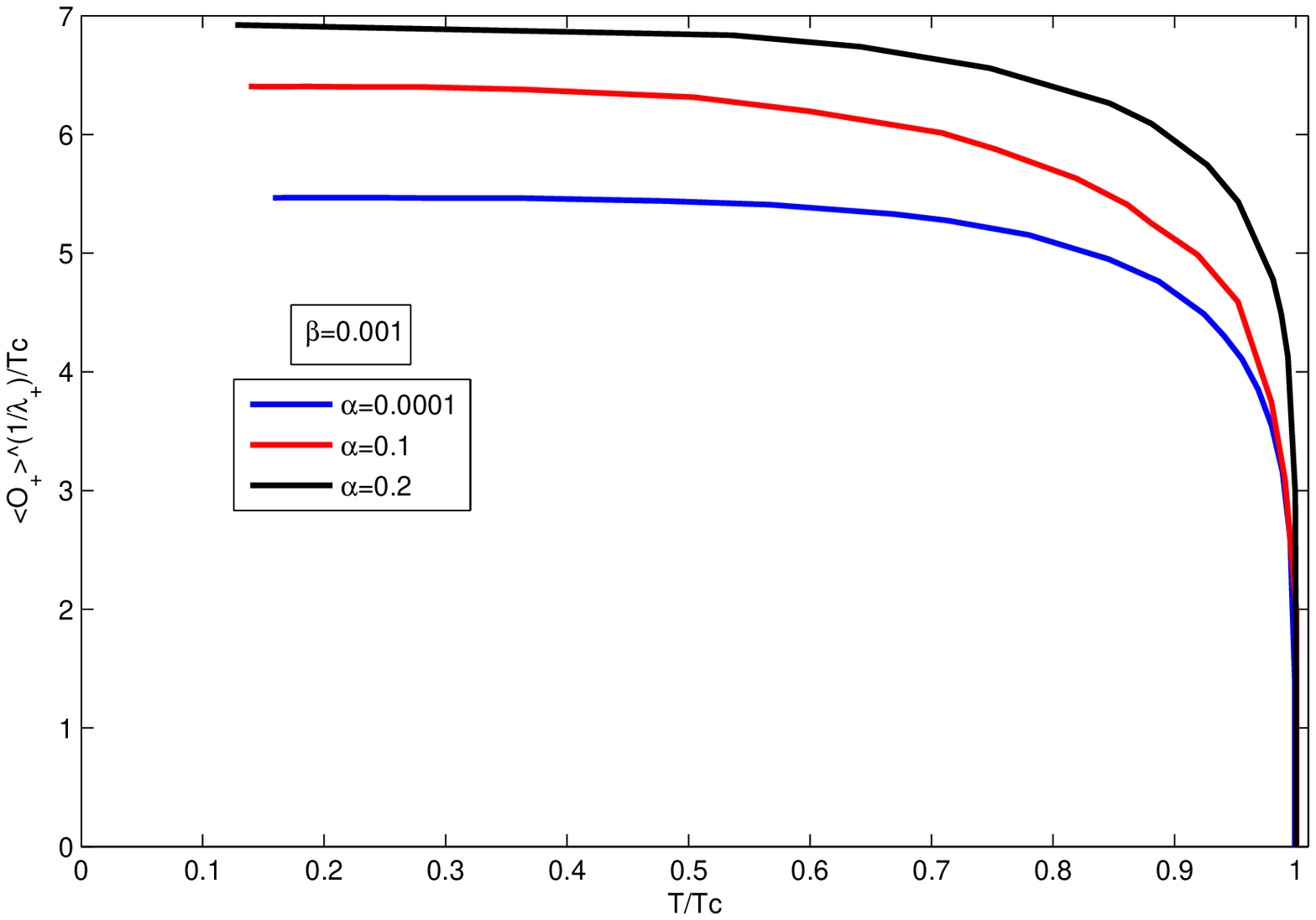}\\ \hspace{0cm}
\caption{The condensation as a function of temperature with
different values of coupling parameters. In the left figure, the
Gauss-Bonnet parameter $\alpha$ is fixed at $0.01$, while the
curvature-cubed constant $\beta$ runs from $0.001$(red line),
$0.01$ (black line) to $0.1$ (blue line) respectively. In the
right figure $\beta$ is fixed at $0.001$ but $\alpha$ varies from
$0.0001$ (blue line) , $0.1$ (red line) to $0.2$ (black line). In
both cases the condensation tends to increase with the coupling
parameters. }}
\end{figure}
Numerically, we obtain the condensation of
$\langle\mathcal{O}_{+}\rangle$ with various positive values of
coupling parameters $\alpha$ and $\beta$. We find that the
critical temperature goes down as either the Gauss-Bonnet constant
$\alpha$ or the curvature-cubed constant $\beta$ increases.
Specifically, if we set $\alpha=0.01$, but change $\beta$ from
0.001 to 0.01 and 0.1, numerical analysis shows that the critical
temperature $T_c$ decreases from $0.1953\rho^{1/3}$ to
$0.1942\rho^{1/3}$ and $0.1739\rho^{1/3}$, respectively. On the
other hand, when $\beta$ is fixed at $0.001$ but $\alpha$
increases from 0.0001 to 0.1 and 0.2, $T_c$ decreases from
$0.1977\rho^{1/3}$ to $0.1755\rho^{1/3}$ and $0.1531\rho^{1/3}$,
which is the same phenomenon as found in Gauss-Bonnet gravity.
Therefore, we conclude that a phase transition occurs in the dual
field theory, but both the curvature-cubed term and the
Gauss-Bonnet term with positive couplings make the condensation
harder in the sense that the critical temperature goes down as the
coupling parameters increase. This conclusion agrees to our
previous analytical results as well. However, it is worthwhile to
point out that $\langle\mathcal {O}_+\rangle^{1/\lambda_+}/T_c$
has larger values when couplings increase.

\section{Electrical Conductivity}
According to the recipe of AdS/CFT dictionary, Maxwell field
$A_\mu$ in the bulk corresponds to a 4-electrical current density
$J_\mu$ on the boundary. In order to calculate the conductivity
$\sigma$ in the boundary theory, we firstly introduce
perturbations of the Maxwell field $\delta A_x$, then consider the
linear response to such perturbations. For simplicity let us
suppose the vector potential is radially symmetric and time
dependent as $\delta A_x(t,r)=A_x(r)e^{-i\omega t}dx$, which
satisfies the EOM as
\begin{eqnarray}\label{A}
A_x^{\prime\prime}+\left(\frac{g^\prime}{g}+\frac{1}{r}\right)A_x^\prime
+\left(\frac{\omega^2}{N^2g^2}-\frac{2\psi^2}{g}\right)A_x =0. \;
\end{eqnarray}
We solve this equation under the incoming boundary condition near
the horizon, namely
\begin{eqnarray}\label{B}
A_x(r) \sim g(r)^{-i\frac{\omega}{4Nr_H}} \ ,
\end{eqnarray}
since this choice gives rise to the retarded Green function of the
system. In the asymptotically AdS region $(r\rightarrow\infty)$,
the general solution behaves as
\begin{eqnarray} \label{C}
A_x=A^{(0)}+\frac{A^{(2)}}{r^2} +\frac{A^{(0)}\omega^2N^4}{2}
\frac{\log\Lambda r}{r^2},
\end{eqnarray}
where $A^{(0)}$, $A^{(2)}$ and $\Lambda$ are integration
constants. As pointed out in \cite{HorRob}, the logarithmic term
in the solution leads to a divergence in the Green function but
this can be removed by introducing an appropriate counter-term in
the action. Taking into account the results obtained in the
previous section and the incoming boundary condition, we can
numerically solve the equation of motion (\ref{A}) and obtain the
values of factors $A^{(0)}$ and $A^{(2)}$. Moreover, the
conductivity can be obtained by means of the retarded Green
function
\begin{eqnarray}\label{D}
  \sigma (\omega) = \frac{1}{i \omega} G^{R}(\omega),
\end{eqnarray}
where $G^R$ can be calculated through the AdS/CFT dictionary as
\cite{Son}
\begin{eqnarray}\label{E}
G^R = - \lim_{r \rightarrow \infty} N g(r) r A_x A_x' .
\end{eqnarray}
Here we have normalized $A_x$ on the boundary. Substituting the
asymptotical solution into the above equations, we can finally
obtain the relation between the conductivity and the frequency as
\begin{eqnarray}\label{F}
\sigma=\frac{2A^{(2)}}{iN\omega A^{(0)}}+\frac{iN^3\omega}{2} \ .
\end{eqnarray}
\begin{figure}
\center{
\includegraphics[scale=0.4]{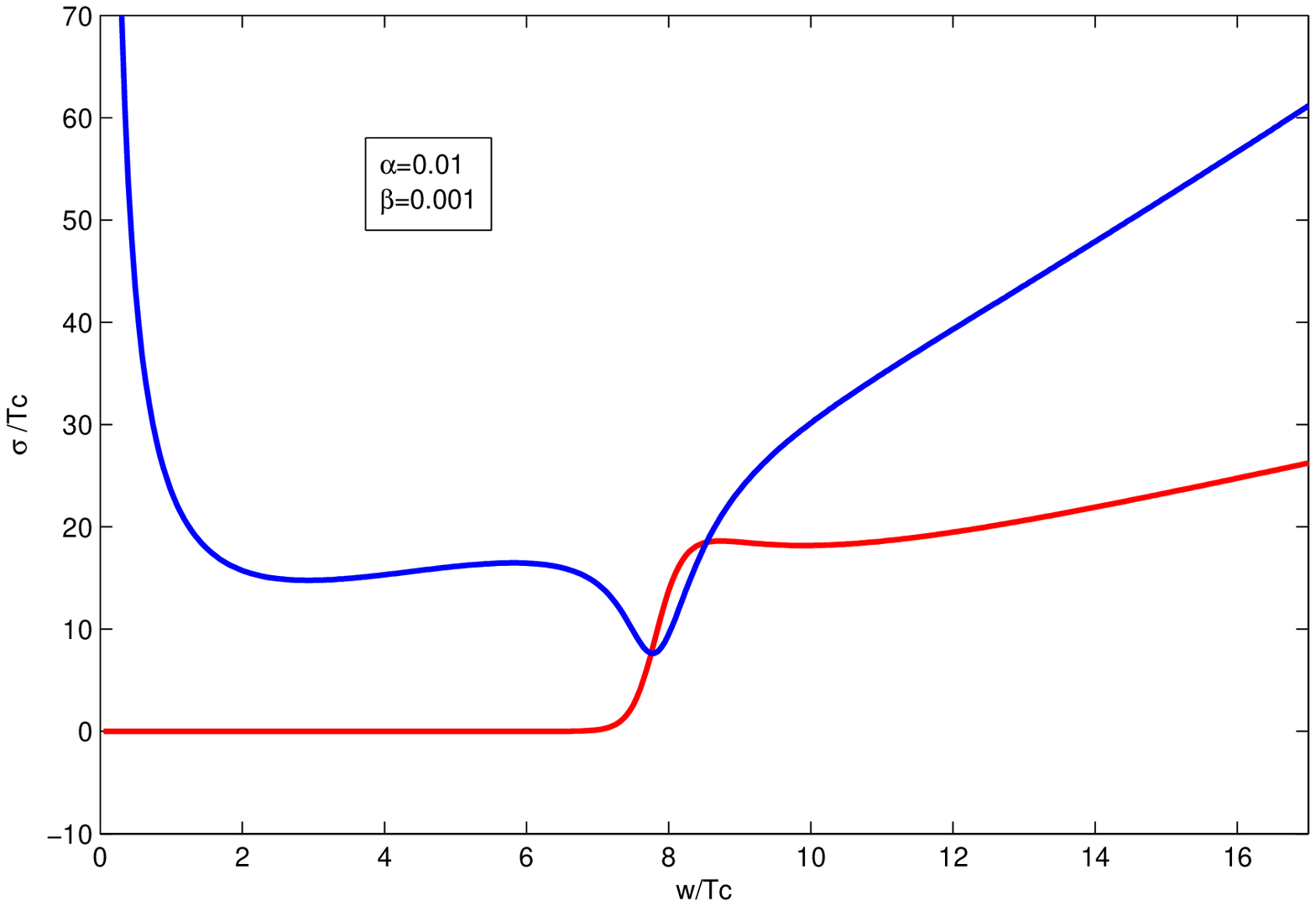}\hspace{0cm}
\includegraphics[scale=0.4]{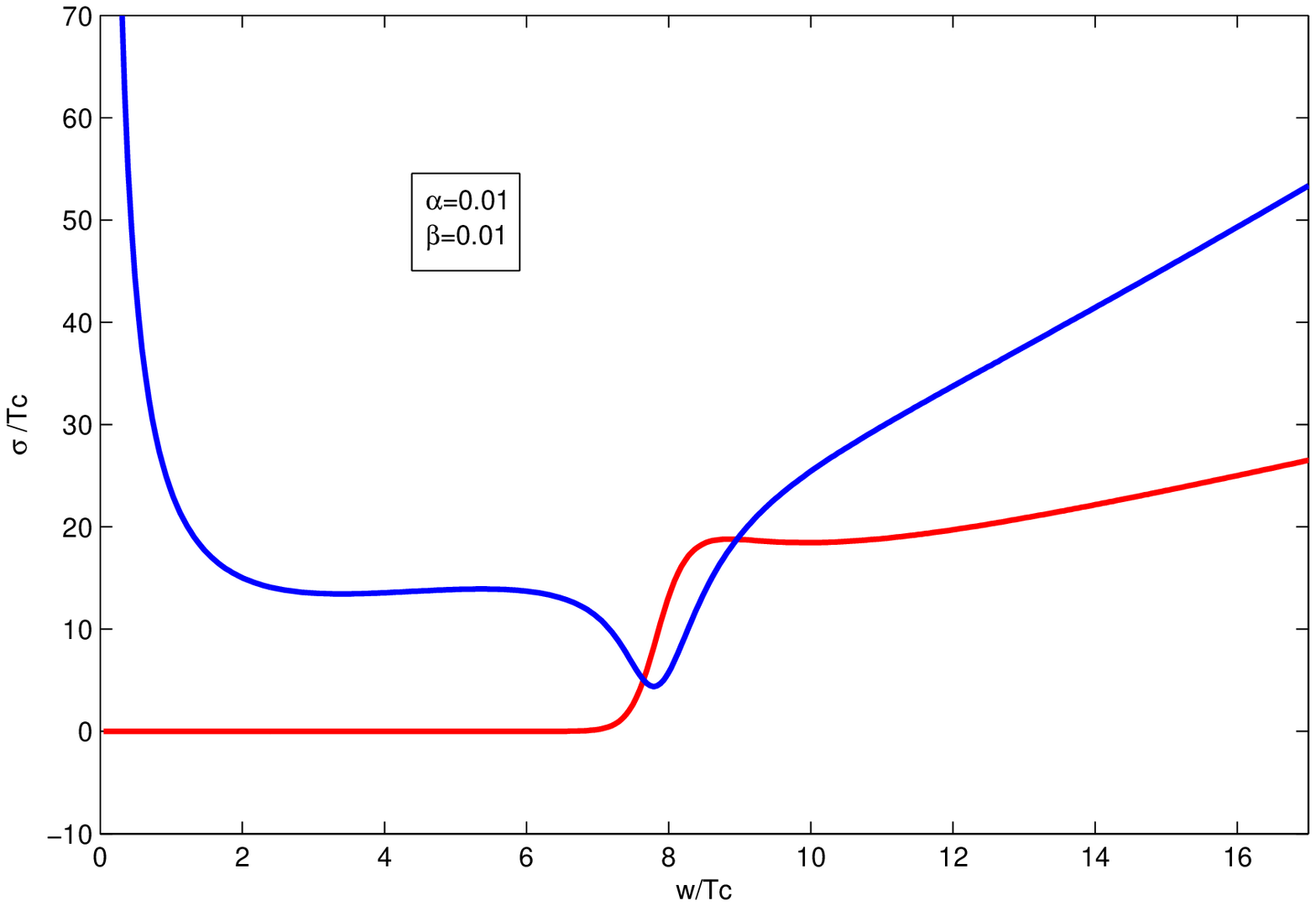}\\ \hspace{0cm}}
\center{
\includegraphics[scale=0.4]{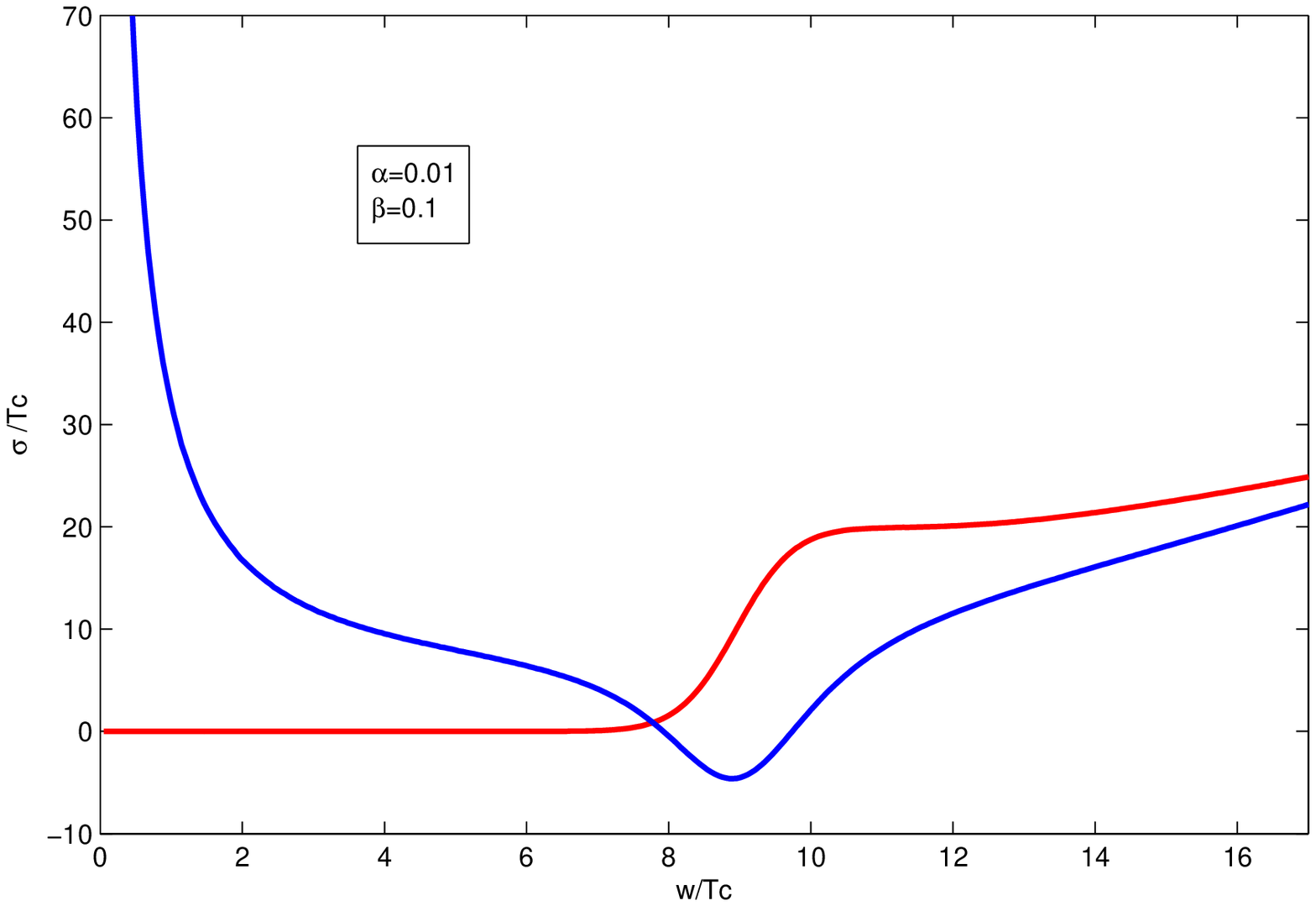}\hspace{0cm}
\includegraphics[scale=0.4]{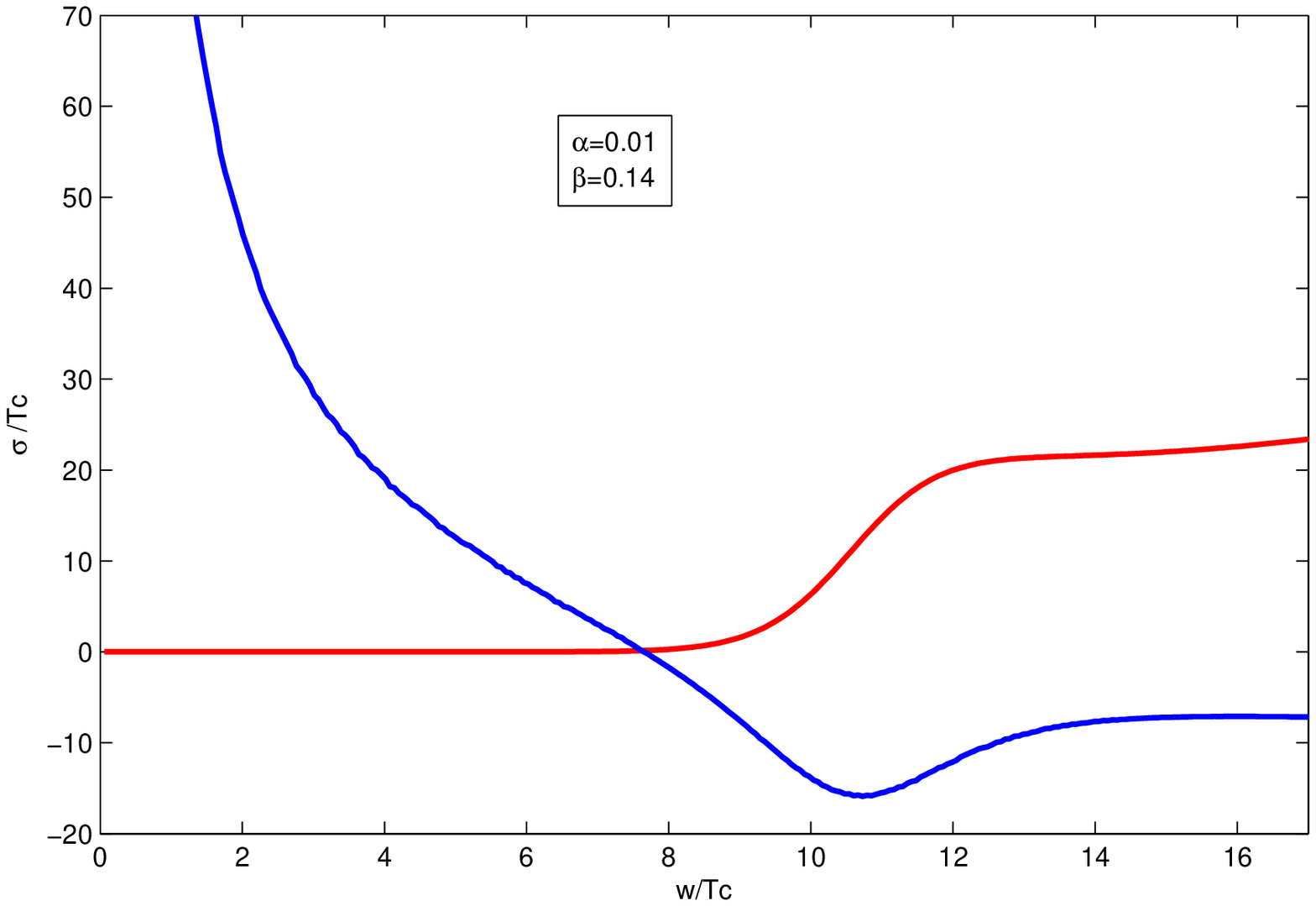}\\ \hspace{0cm}
\caption{Conductivity for superconductors with fixed $\alpha=0.01$
for different values of $\beta$ at the same temperature
$T/T_c=0.153$.}}
\end{figure}
The numerical results with different couplings are illustrated in
FIG. 5 and FIG. 6, where red lines and blue lines represent the
real part and imaginary part of $\sigma$, respectively. In FIG. 5,
we fix $\alpha=0.01$ but set $\beta$ to be $0.001$, $0.01$, $0.1$
and $0.14$, respectively. All these lines are plotted at
temperature $T/T_c=0.153$. On the other hand, in FIG. 6 we fix
$\beta=0.001$ but change $\alpha$ from $0.0001$, $0.1$, $0.2$ to
$0.245$ at temperature $T/T_c=0.158$.
\begin{figure}
\center{
\includegraphics[scale=0.4]{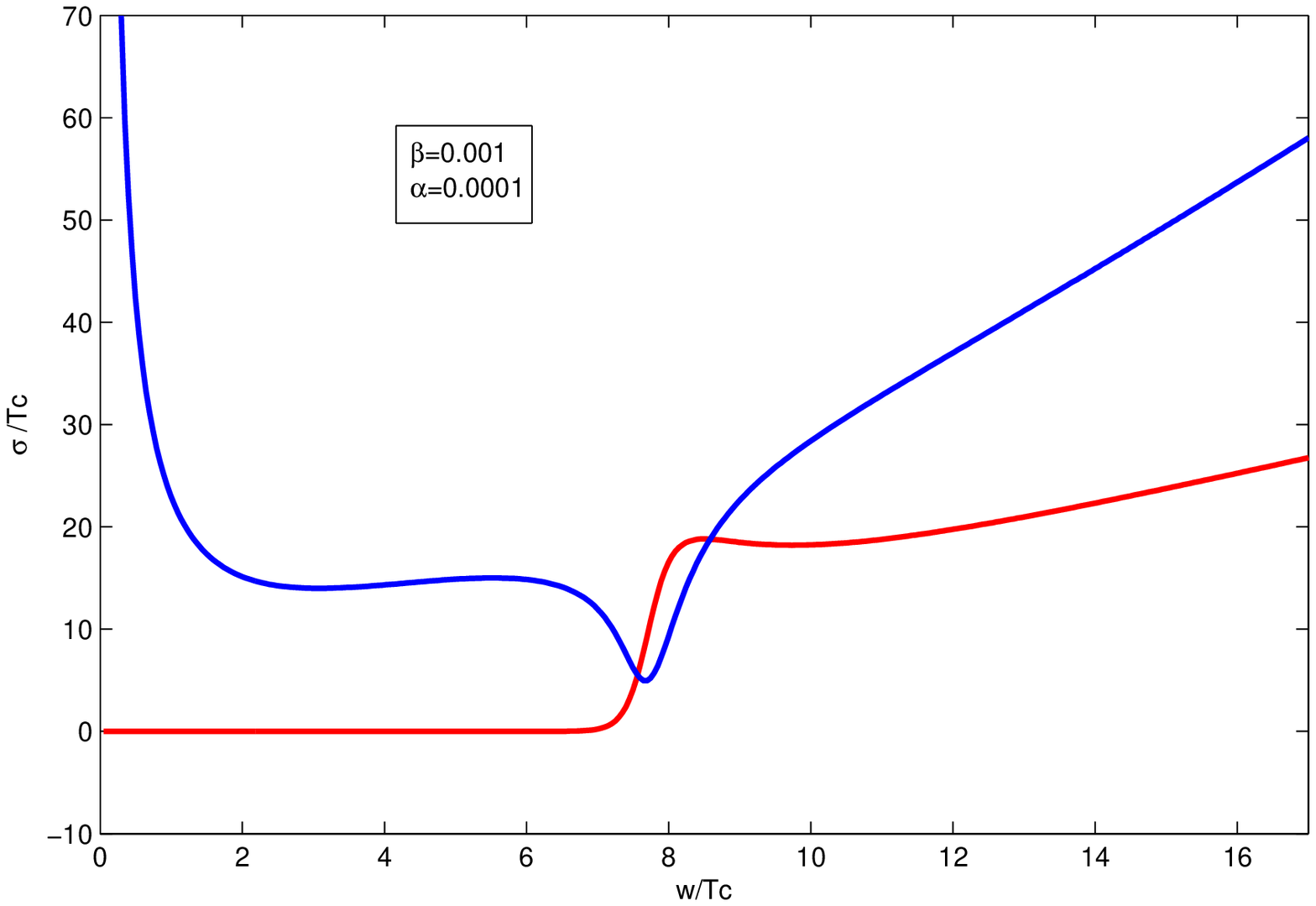}\hspace{0cm}
\includegraphics[scale=0.4]{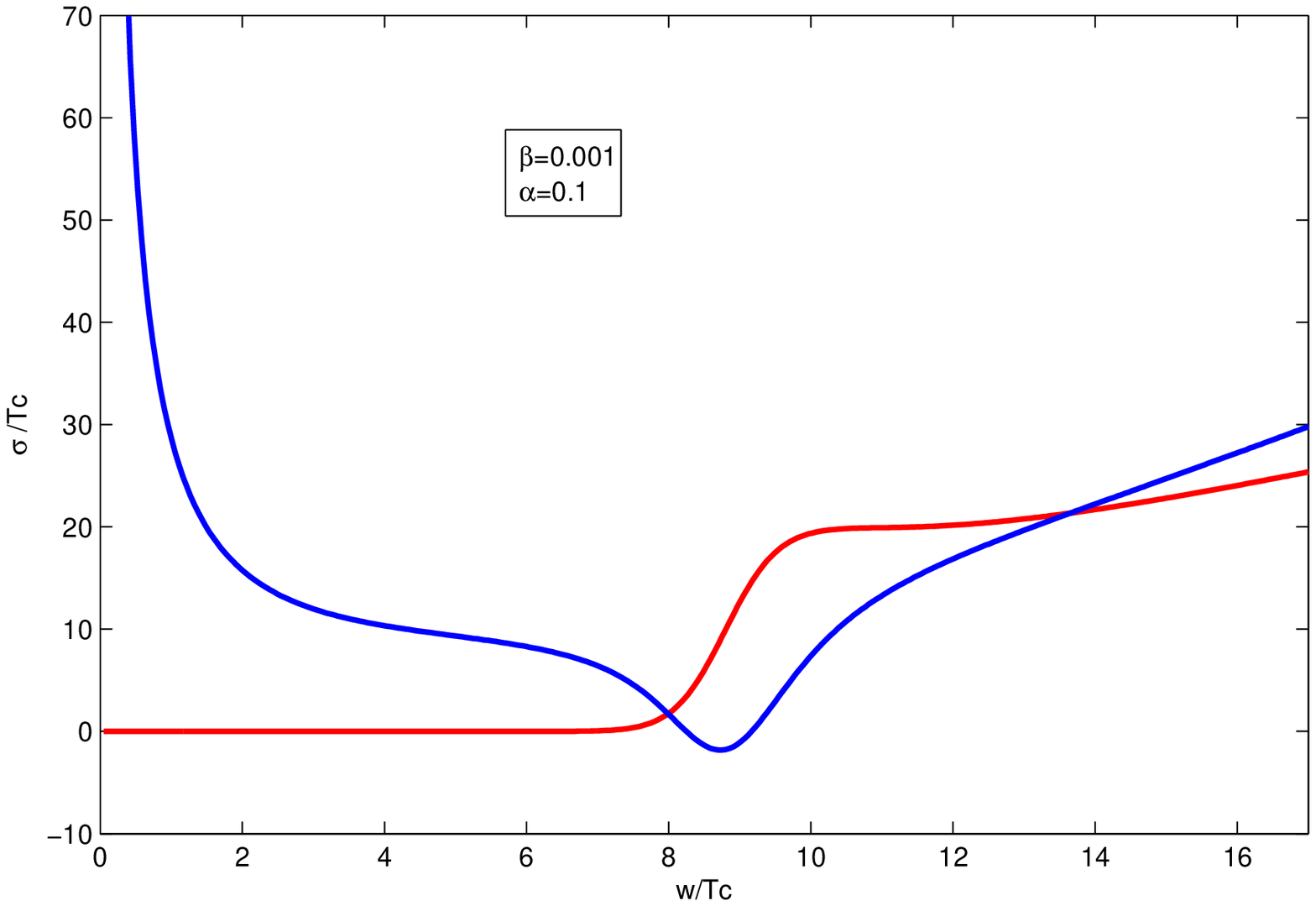}\\ \hspace{0cm}}
\center{
\includegraphics[scale=0.4]{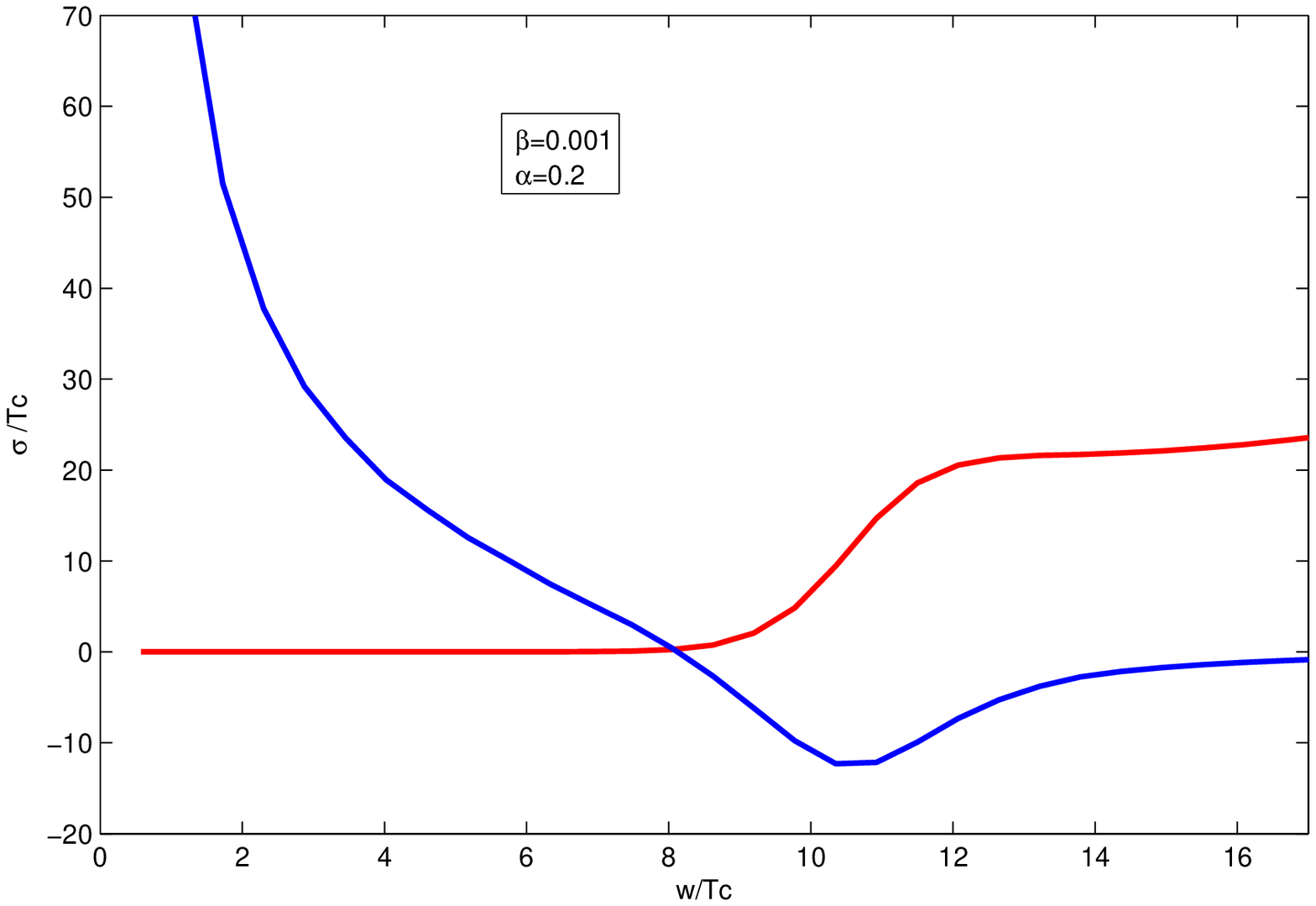}\hspace{0cm}
\includegraphics[scale=0.4]{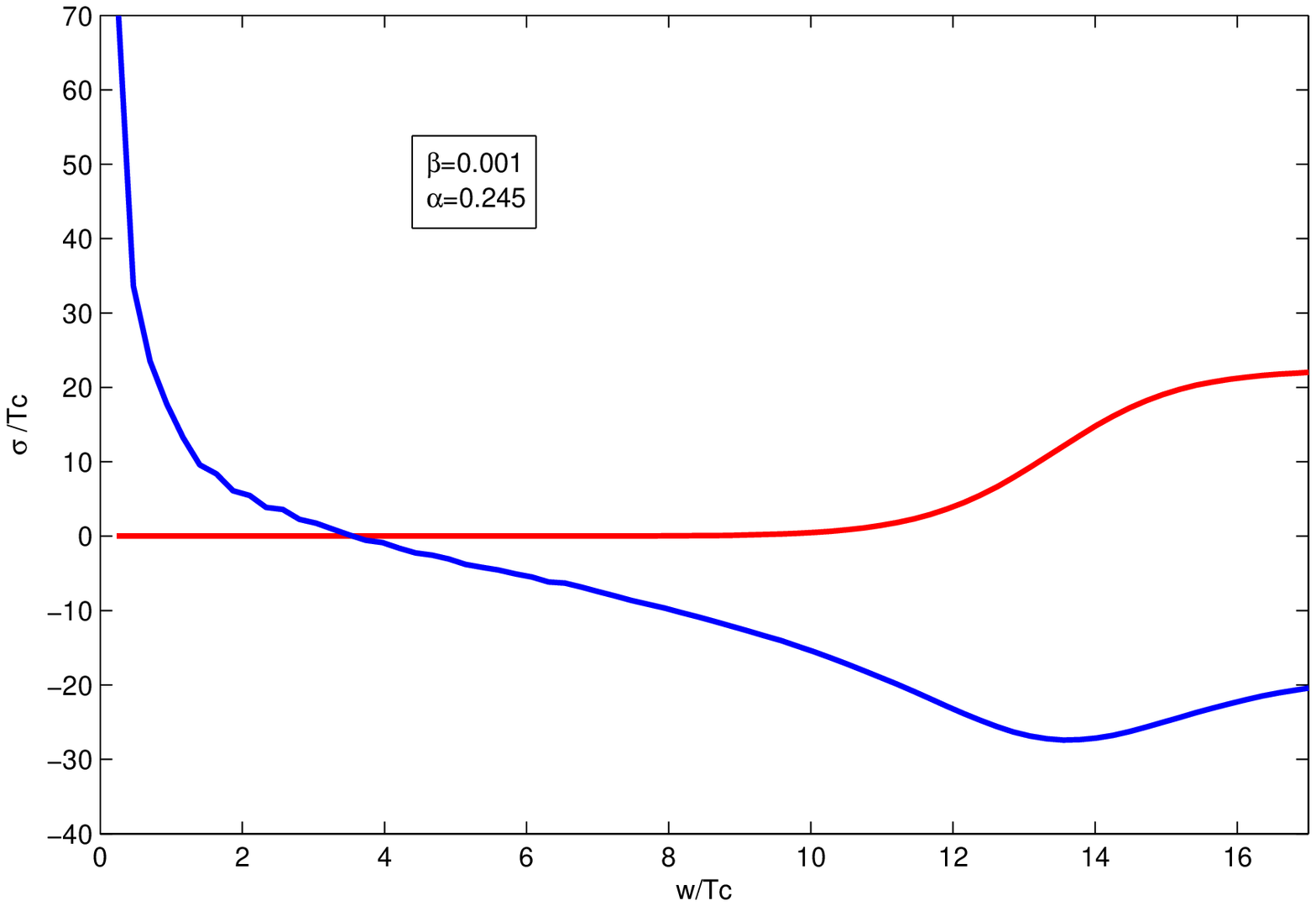}\\ \hspace{0cm}
 \caption{Conductivity for superconductors with fixed
$\beta=0.001$ for different values of $\alpha$ at the same
temperature $T/T_c=0.158$.}}
\end{figure}
From the real part of the conductivity, we notice a gap in the
conductivity with the frequency $\omega_g$. In Einstein's case, it
has been found that there is a remarkable universal relation
$\frac{\omega_g}{T_c} \simeq 8$ for various condensations
\cite{HorRob}. This ratio is more than twice of
 the value in weakly coupled BCS theory, implying that the
holographic superconductor is strongly coupled. In our case as
both coupling parameters are small, the relation
$\frac{\omega_g}{T_c} \simeq 8$ holds indeed. However, when
quadratic and cubic corrections become strong,
$\frac{\omega_g}{T_c}$ is no longer fixed but running with the
values of coupling parameters. From these figures, we easily
notice that this ratio enlarges with the increase of $\alpha$ or
$\beta$, until it reaches a maximal value when coupling parameters
take values  on the border line as shown in FIG.1. This is quite
similar to what happens in GB gravity\cite{Gregory}. But different
from GB theory, the cubic term will intensively suppress the
imaginary part of the conductivity when $\beta$ is large enough.
For the imaginary part of the conductivity, there also exists a
pole at $\omega=0$, indicating that the real part of the
conductivity contains a delta function according to the
Kramers-Kronig relation.
\section{General holographic superconductors with non-positive couplings}
In previous sections we have studied the holographic
superconductivity for the case that both coupling parameters
$\alpha$, $\beta$ are positive. In this section we intend to study
the holographic superconductor in quasi-topological gravity with
non-positive couplings in a parallel way. The main difference is
that for these coupling parameters, one need consider different
kinds of AdS black hole backgrounds, which comes from the fact
that there exist three distinct solutions to Eq.(\ref{c}). Besides
$f_3$, the other two solutions are
\begin{eqnarray}
f_1(r)&=&u+v-\frac{\alpha}{3\beta},\label{H}\\
f_2(r)&=&-\frac{1}{2}(u+v)+i\frac{\sqrt{3}}{2}(u-v)-\frac{\alpha}{3\beta}
,\label{G}
\end{eqnarray}
where $u$ and $v$ are defined in Eq.(\ref{k}). As illustrated in
figure one in \cite{Myers}, for different regions in parameter
space, we need choose the appropriate solution to obtain stable
AdS background. Next we focus our discussion on the cases that
$\alpha$ and $\beta$ are not both positive. Without loss of
generality, we may pick up some points in different regions of the
parameter space and obtain the corresponding critical temperatures
numerically. The results are summarized as TABLE \ref{table1}.
\begin{widetext}
\begin{center}
\begin{table} [h!]
\begin{tabular} { |c|c|c|c|c|c|c|c|c|c| }
\hline
& \multicolumn{3}{|c|}{$f_1$} & \multicolumn{3}{|c|}{$f_2$} & \multicolumn{3}{|c|}{$f_3$}
 \\
\hline
 \hline
$\alpha$
&$-0.25$&$-0.1$&$-0.05$&$0.24$&$0.12$&$0.04$&$-0.10$&$-0.10$&$-0.35$
          \\
        \hline
$\beta$
&$-0.02$&$-0.0002$&$-0.0002$&$-0.0020$&$-0.0020$&$-0.0004$&$0.1000$&$0.0005$&$0.1000$
          \\
        \hline
$T_{c}/\rho^{1/3}$&
$0.2438$&$0.2152$&$0.2067$&$0.1384$&$0.1750$&$0.1908$&$0.1986$&$0.2151$&$0.2446$
          \\
 \hline
\end{tabular}
\caption{The dependence of the critical temperature on two
couplings for different black hole backgrounds.} \label{table1}
\end{table}
\end{center}
\end{widetext}
From this table, one finds a general rule that for smaller coupling
parameters the value of the critical temperature becomes larger.
More explicitly, as one of the parameters is fixed, we find the
condensation is always becoming harder with the increase of the
other parameter. This observation is consistent with our results
obtained in previous sections. As a result, when the coupling
parameters take negative values, it is possible to construct a
superconductor model with a critical temperature below the one in
Einstein's gravity theory, which is $T_{co}\equiv 0.198\rho^{1/3}$.
As a matter of fact, when both parameters are negative, our
numerical results show that the critical temperature is always above
$0.198\rho^{1/3}$, which is in contrast to our previous results
where the temperature is always below this value for both positive
parameters. A more delicate situation occurs when one parameter is
positive while the other is negative. Whether the critical
temperature is above or below $T_{co}$ depends on the competition of
the fluctuation effects due to these two couplings. One special case
is that these two effects are cancelled out such that the critical
temperature is equal to $T_{co}$. For explicitness we plot the
trajectory of $T_c=T_{co}$ numerically in the parameter space and
the final result is illustrated in FIG.7. In this figure it is
manifest that in the white region $T_c<T_{co}$ such that the
condensation is suppressed, while in the purple region $T_c>T_{co}$
and the condensation is easier comparing with that in Einstein's
gravity theory.
\begin{figure} \center{
\includegraphics[scale=0.4]{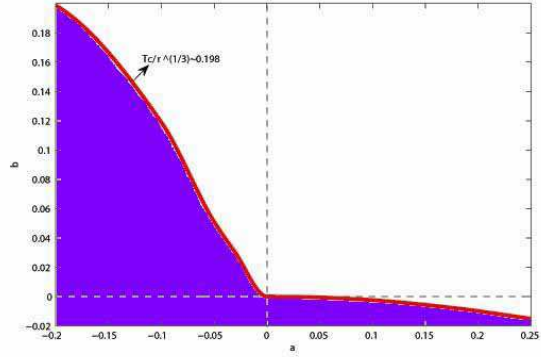}\\ \hspace{0.0cm}
\caption{The trajectory of $T_c=T_{co}=0.198\rho^{1/3}$ in
parameter space. Comparing with that in Einstein's theory, the
condensation becomes harder in the white region but easier in the
purple region.}}
\end{figure}
\begin{figure} \center{
\includegraphics[scale=0.26]{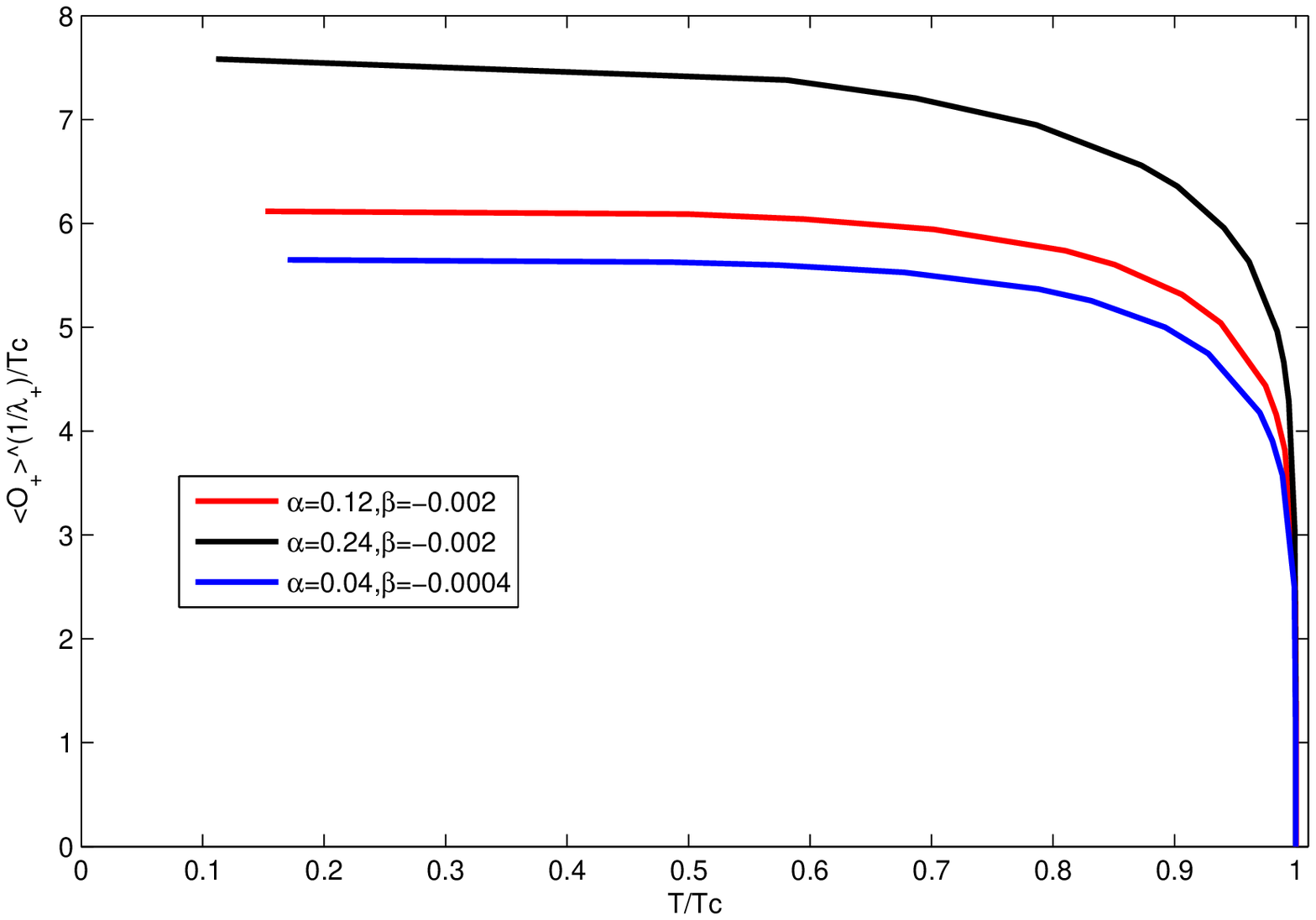}\hspace{0.0cm}
\includegraphics[scale=0.26]{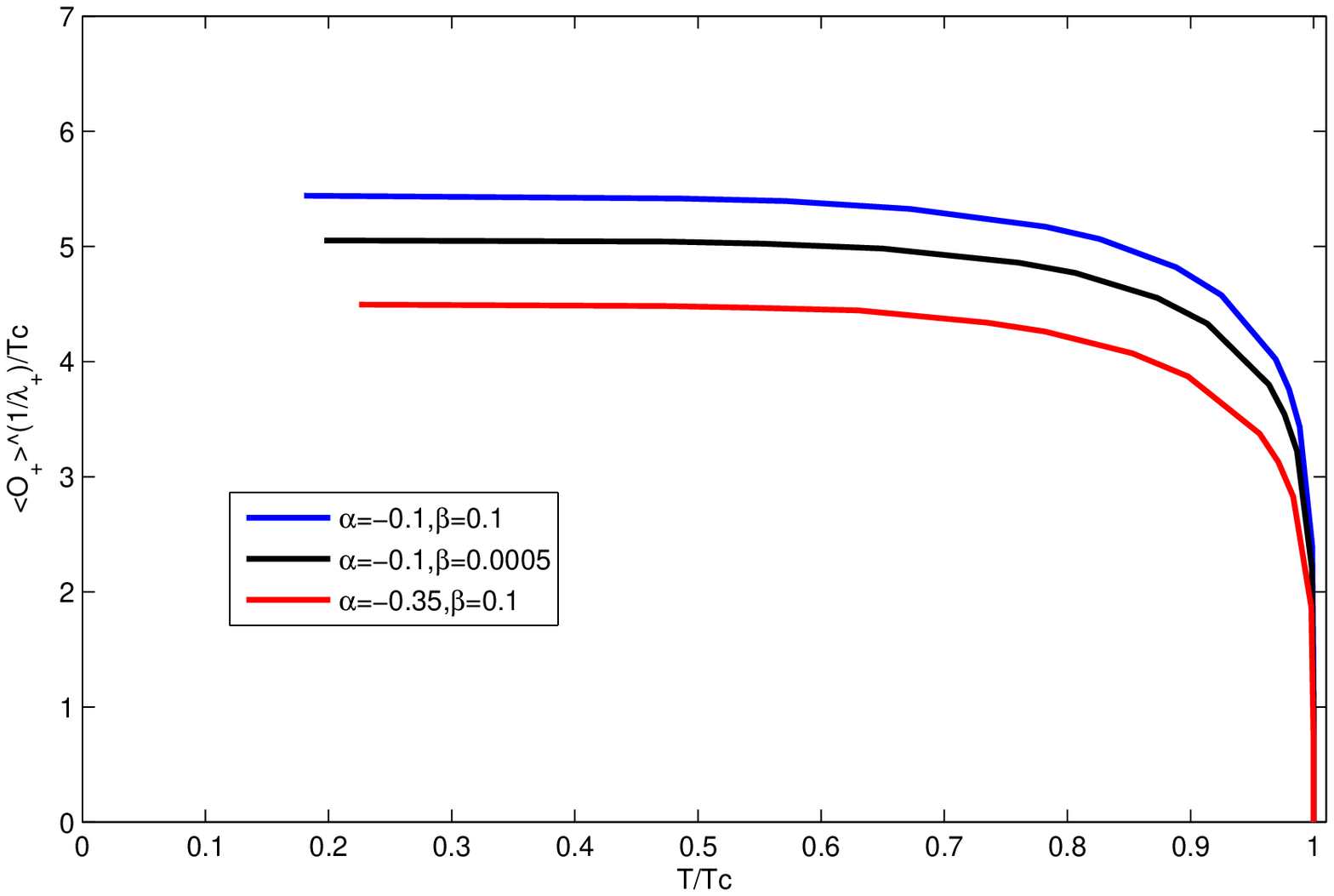} \hspace{0.0cm}
\includegraphics[scale=0.26]{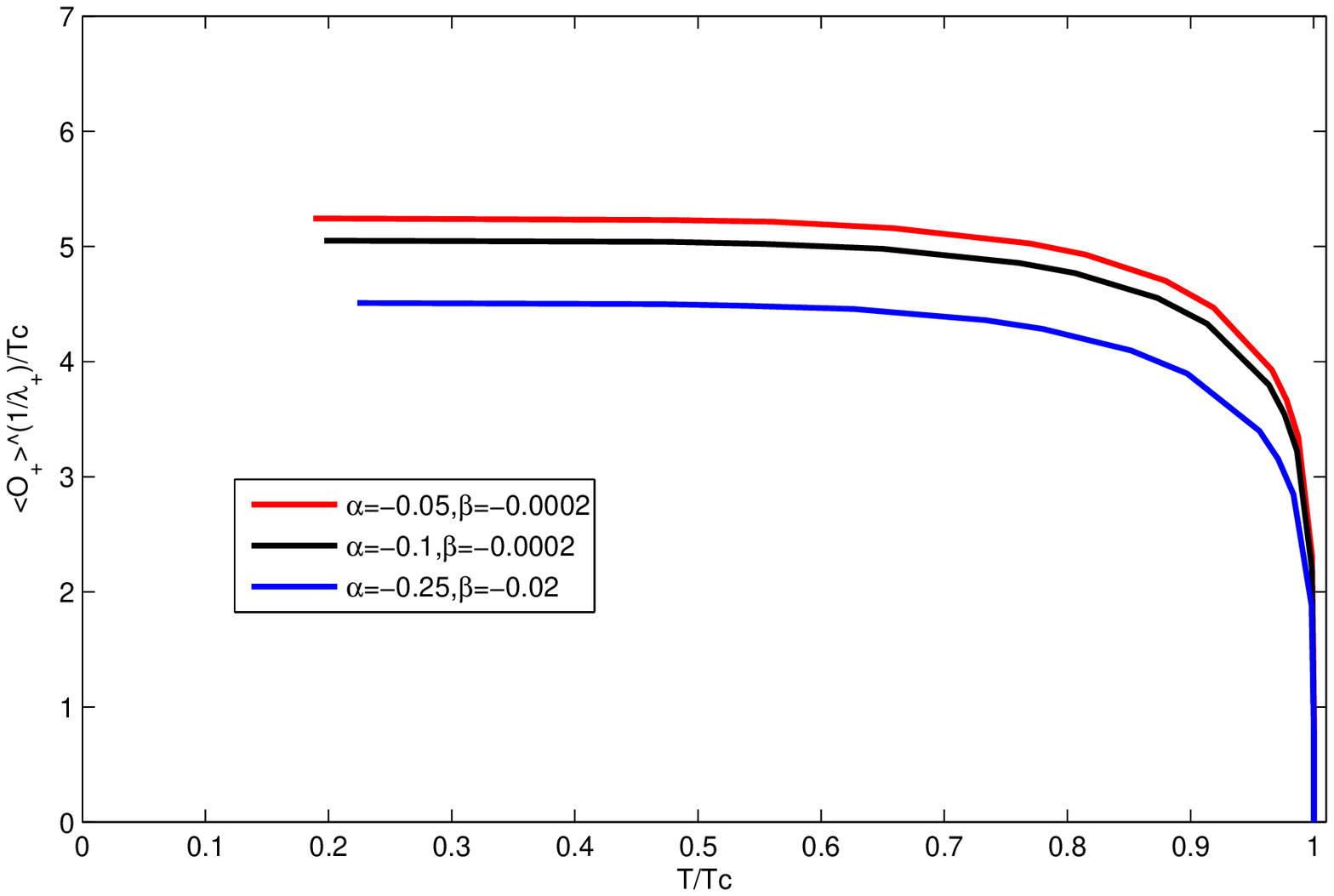}\\ \hspace{0.0cm}
\caption{The condensations for $\alpha>0$ and $\beta<0$ in the left
figure, $\alpha<0$ and $\beta>0$ in the middle figure, and
$\alpha<0$ and $\beta<0$ in the right one. Notice that the
corresponding black hole backgrounds are specified by $f_2$, $f_3$
and $f_1$, respectively. }}
\end{figure}
\begin{figure} \center{
\includegraphics[scale=0.26]{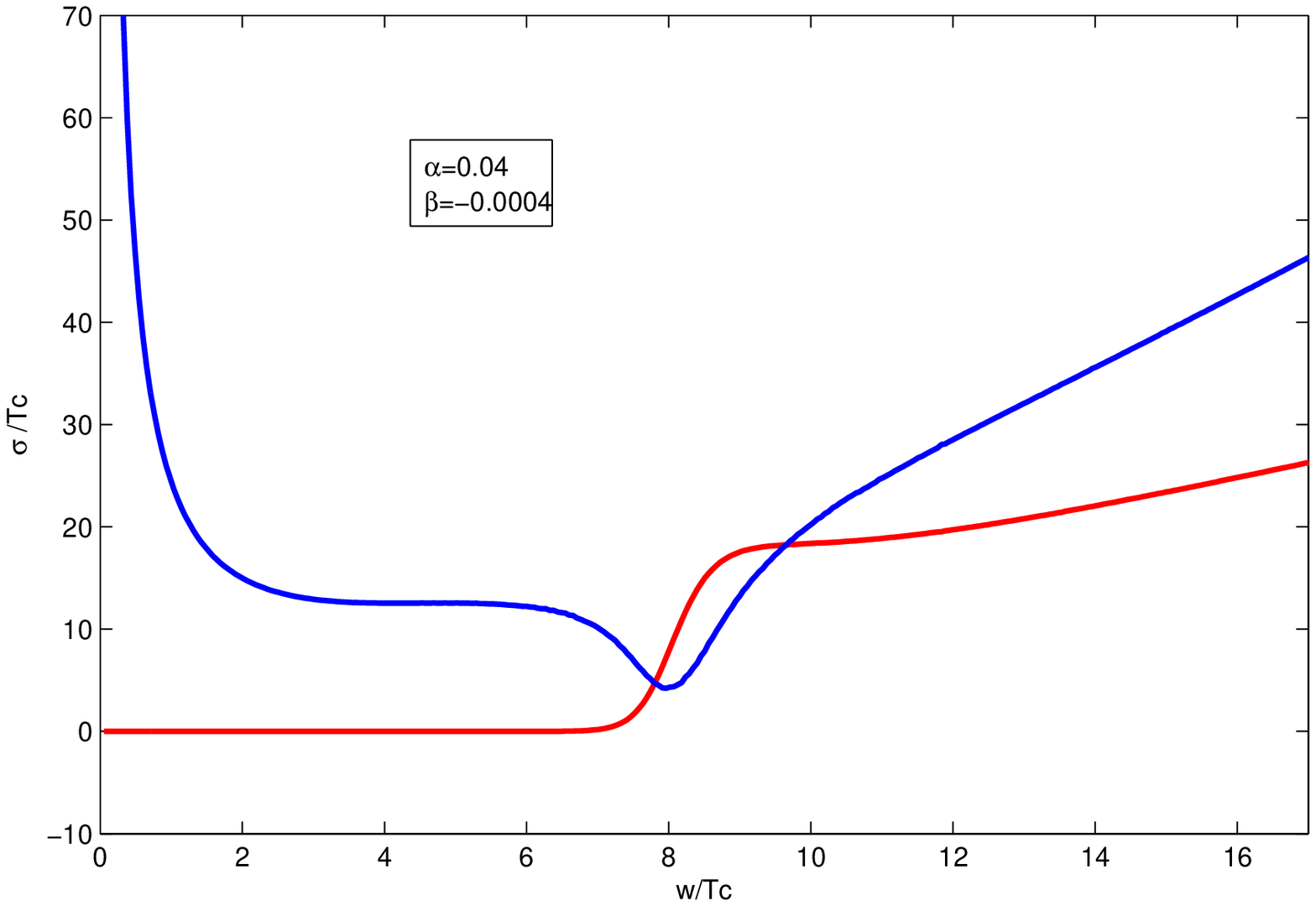}\hspace{0cm}
\includegraphics[scale=0.26]{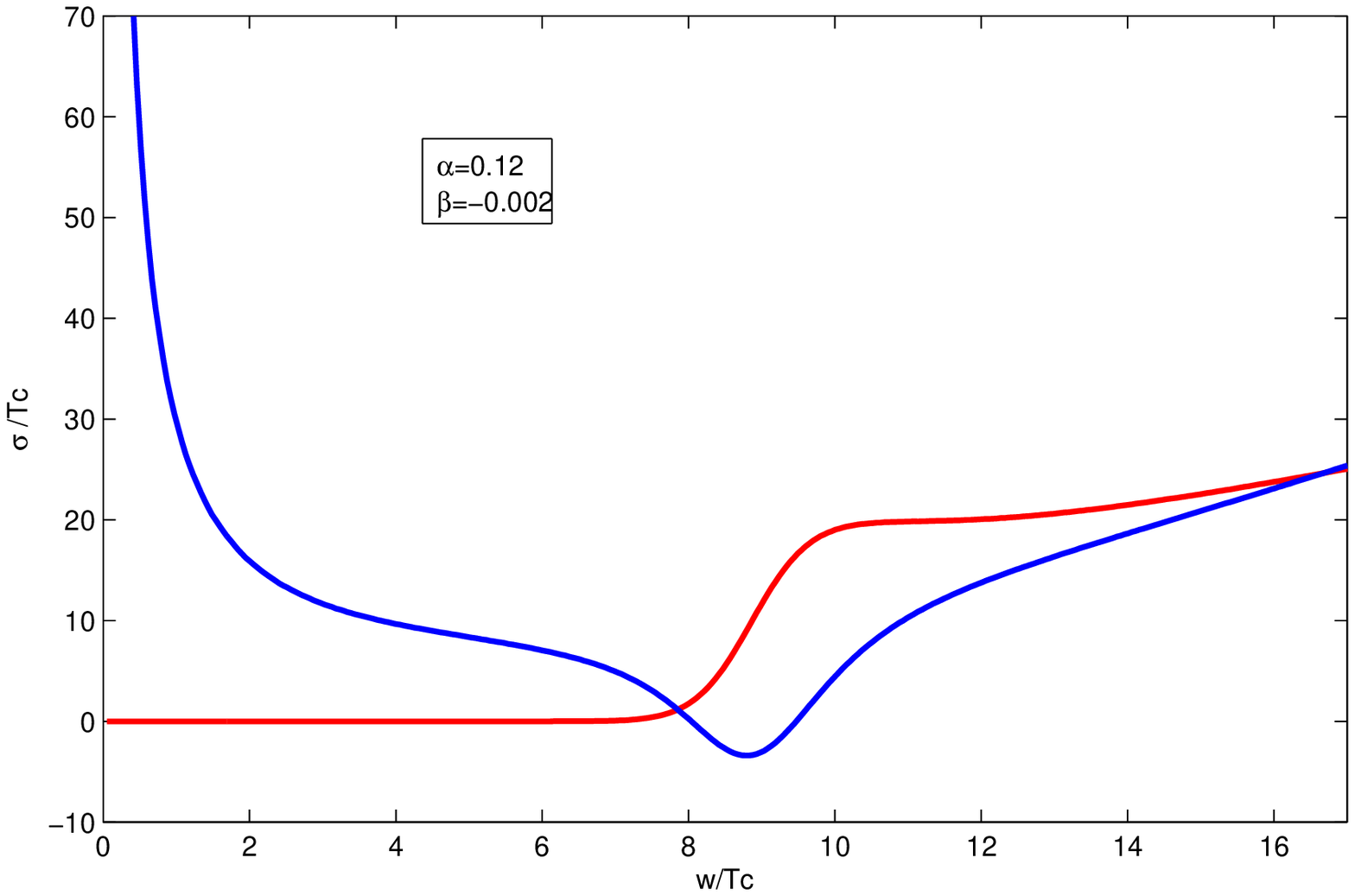} \hspace{0cm}
\includegraphics[scale=0.26]{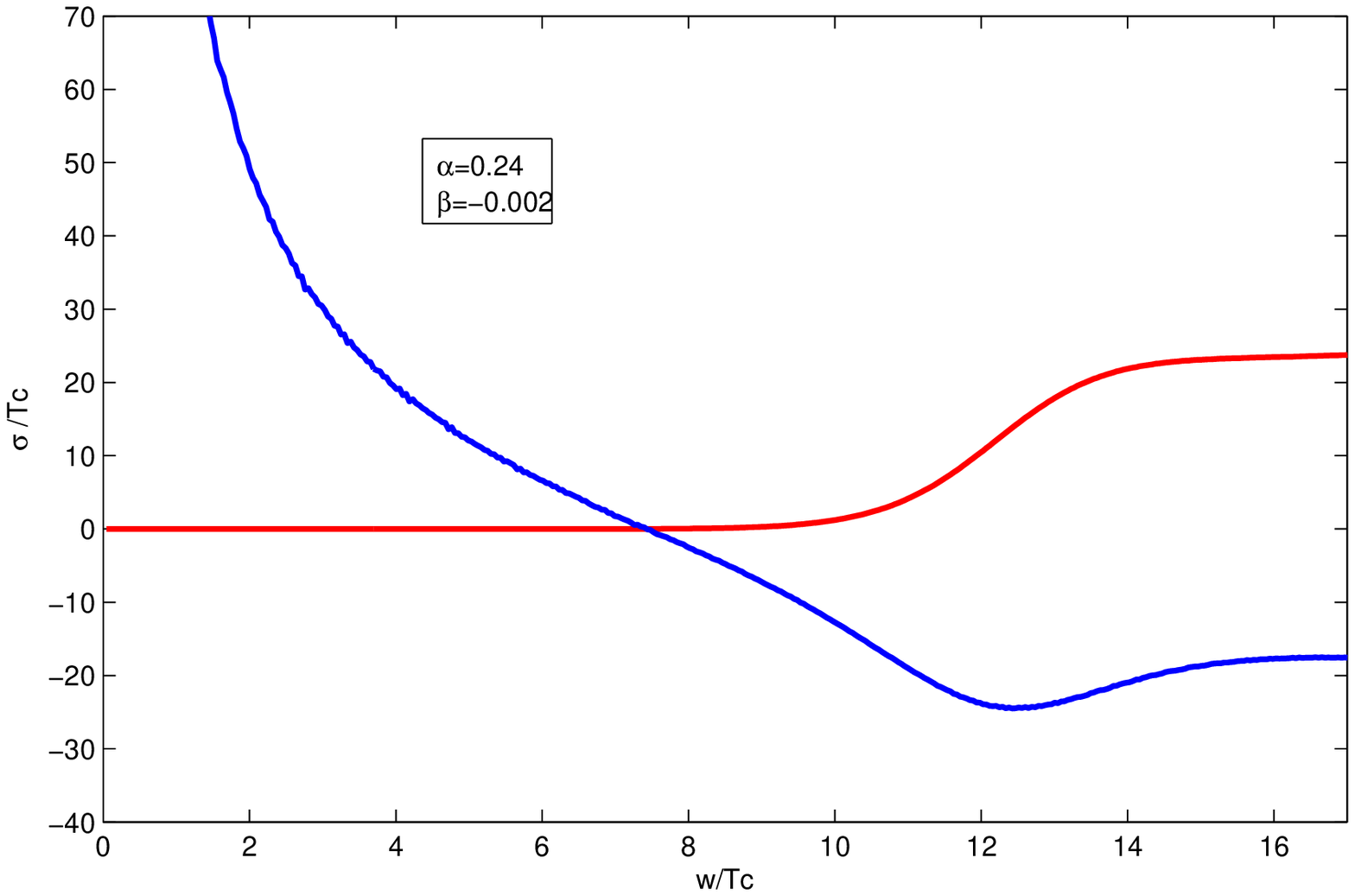}\\ \hspace{0cm}
\caption{The conductivities for $\alpha>0$ and $\beta<0$. With the increase of $\alpha$ and the decrease of $\beta$,
the value of $\omega_g/T_c$ runs from about 8 to more than 12. }}
\end{figure}
\begin{figure} \center{
\includegraphics[scale=0.26]{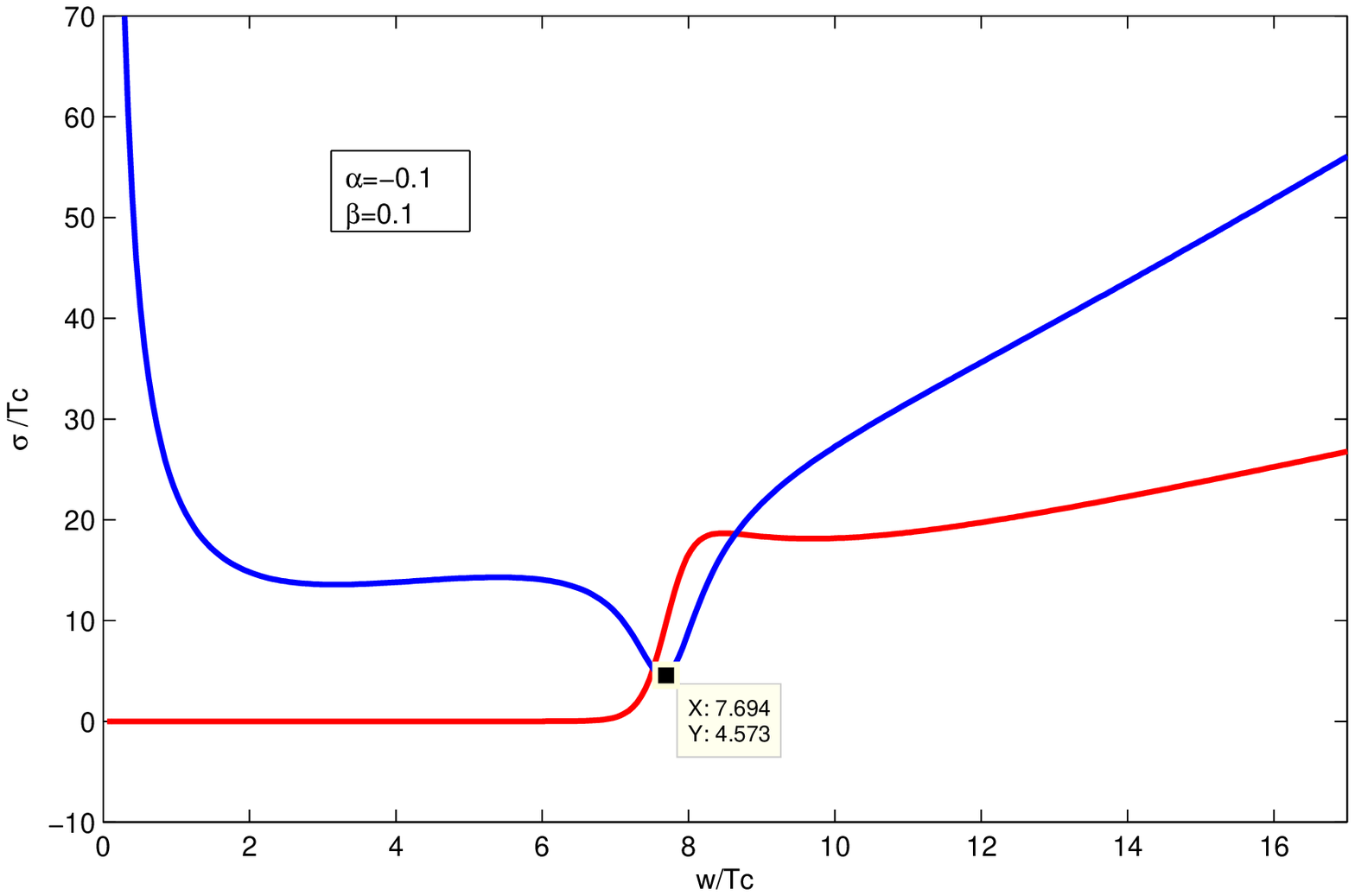} \hspace{0cm}
\includegraphics[scale=0.26]{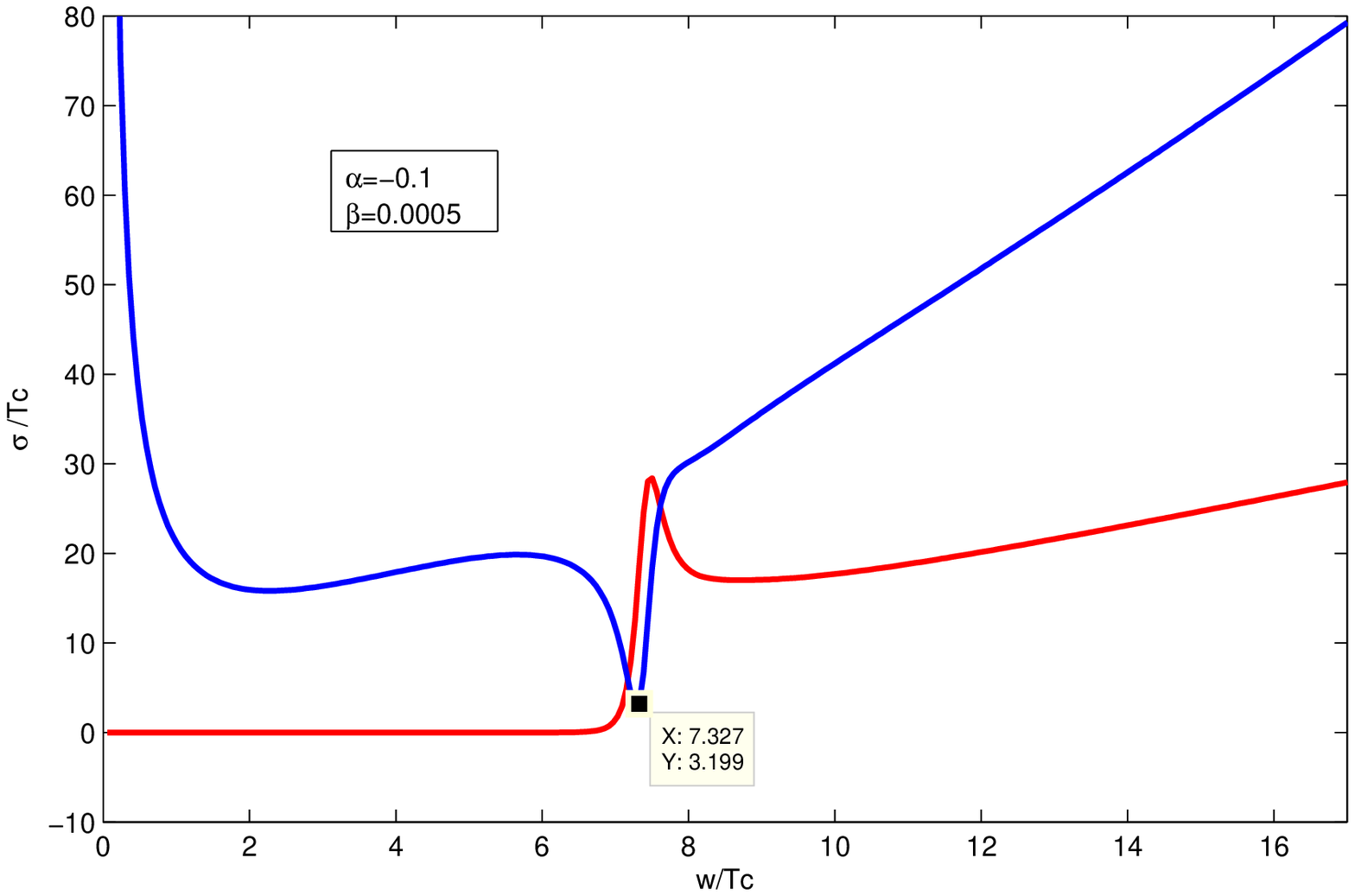} \hspace{0cm}
\includegraphics[scale=0.26]{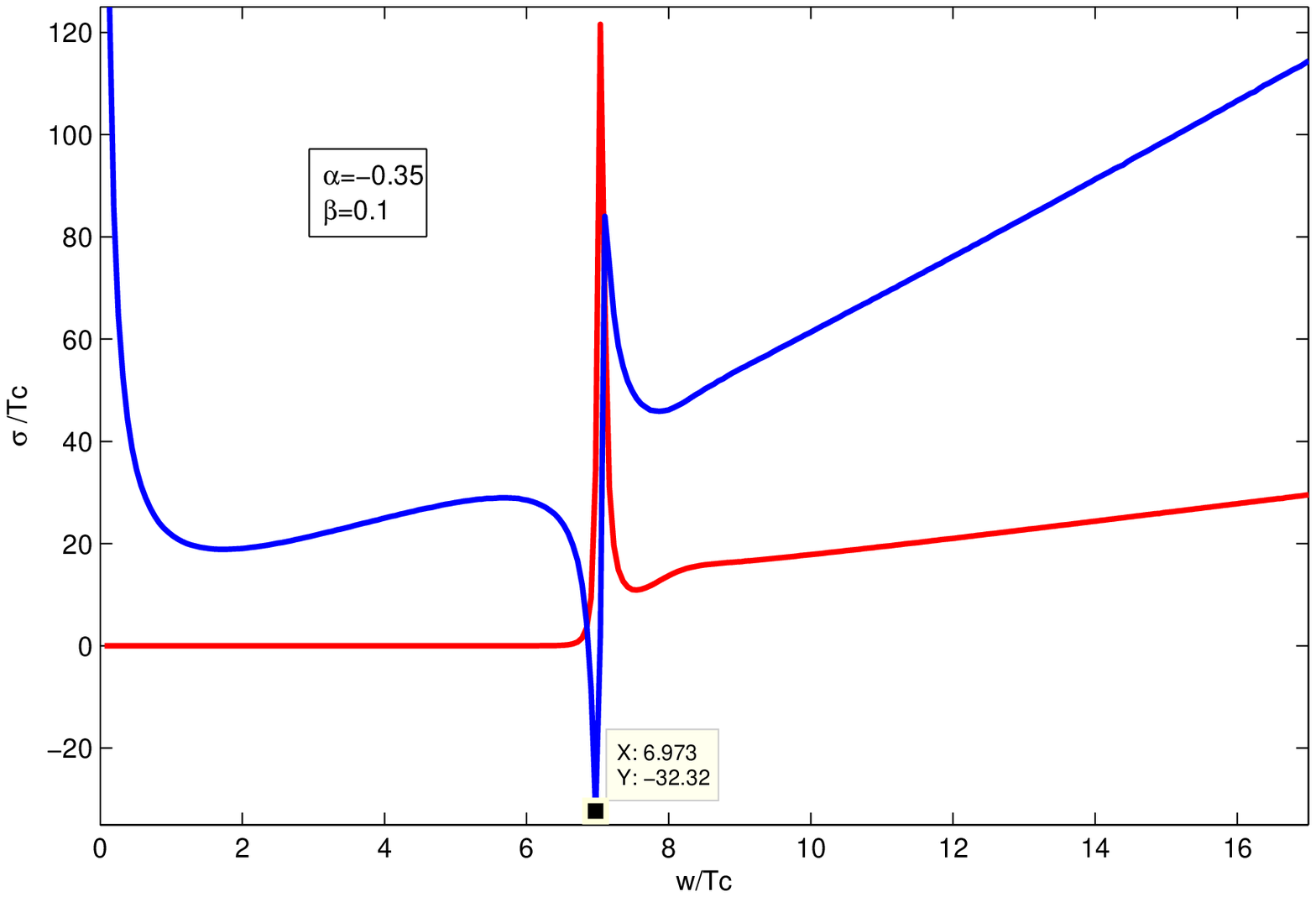}\\ \hspace{0cm}
\caption{The conductivities for $\alpha<0$ and $\beta>0$. A spike exists in the rightmost figure. }}
\end{figure}
\begin{figure} \center{
\includegraphics[scale=0.26]{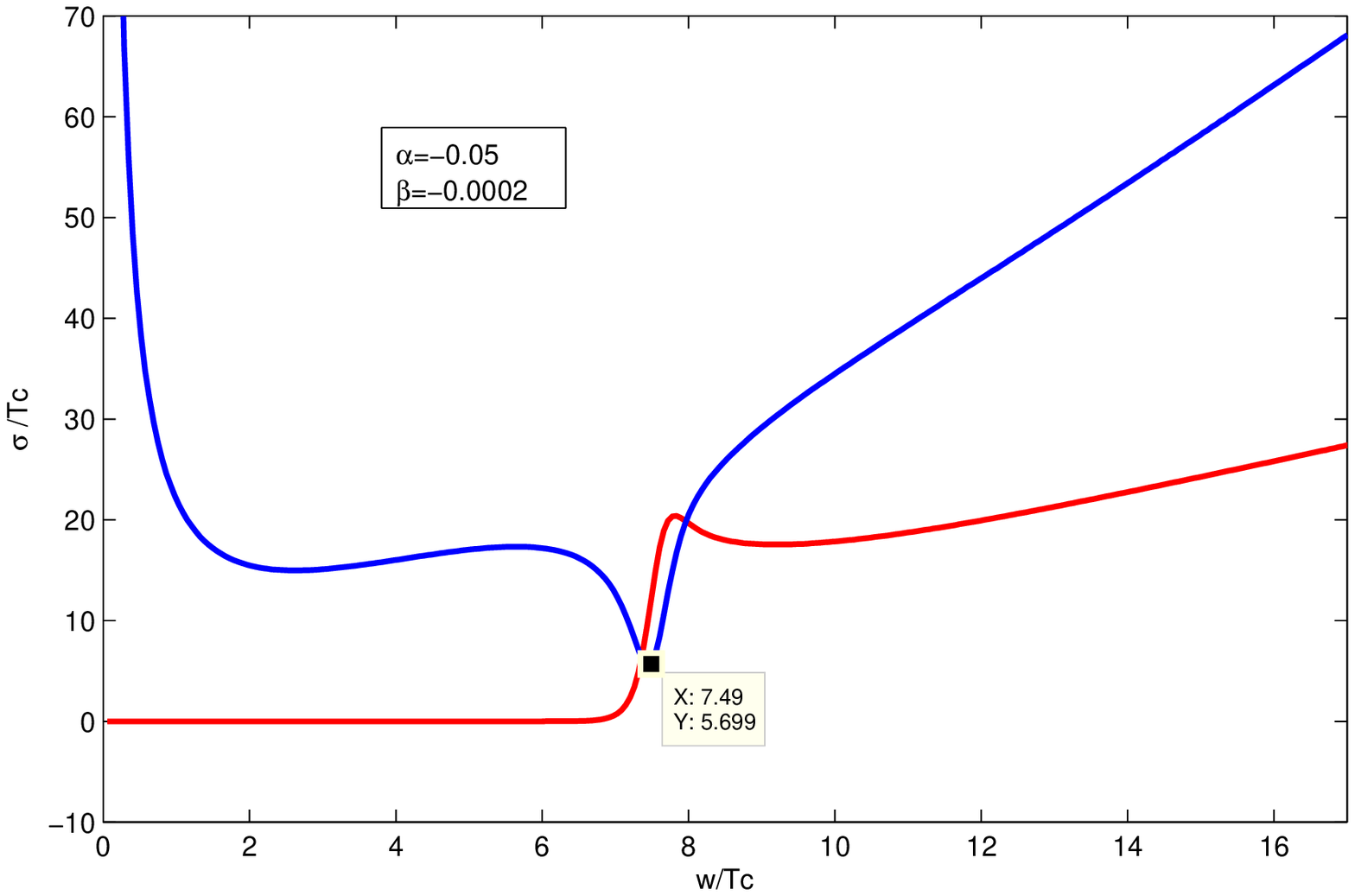}\hspace{0cm}
\includegraphics[scale=0.26]{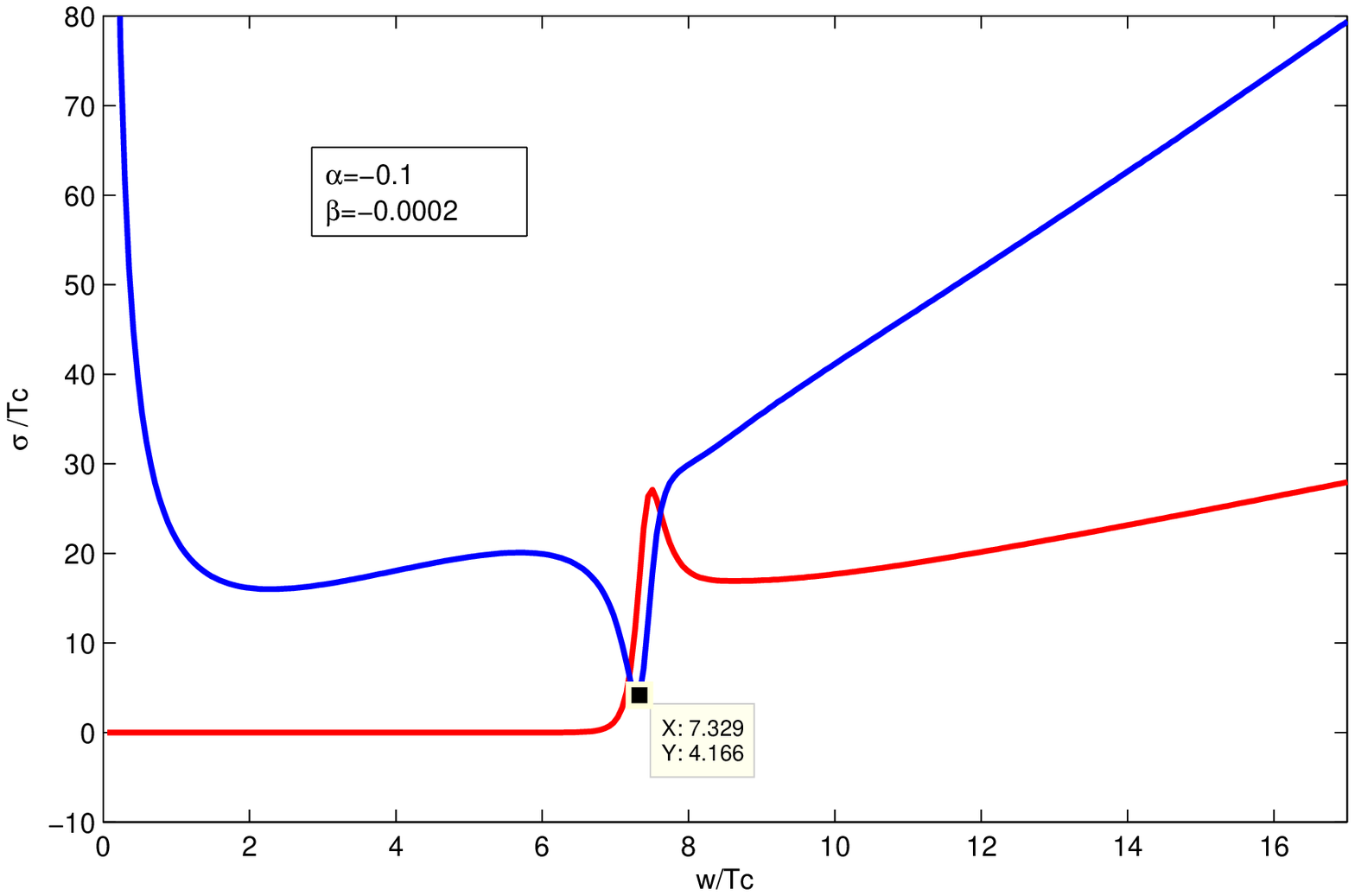} \hspace{0cm}
\includegraphics[scale=0.26]{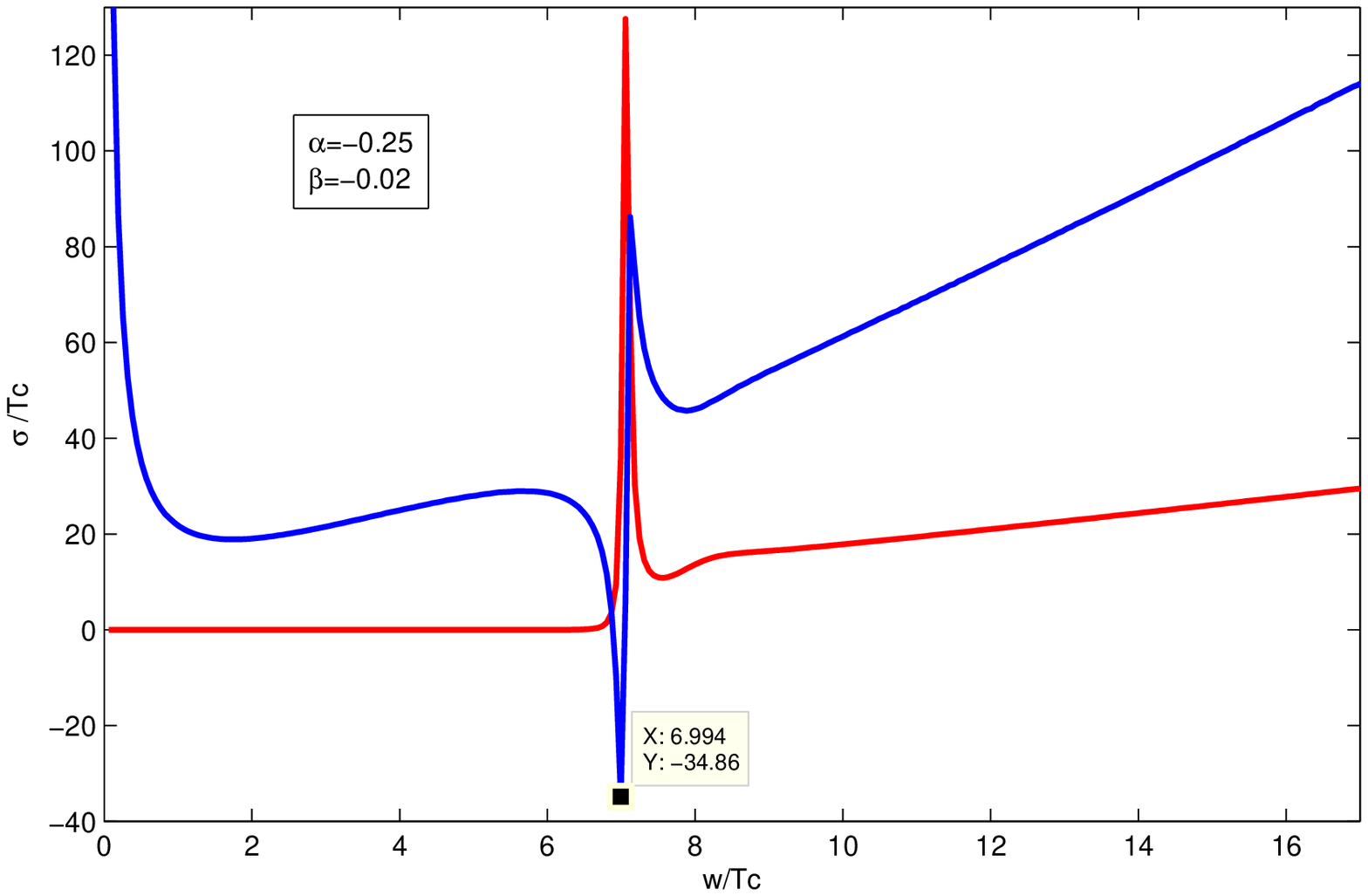}\\ \hspace{0cm}
\caption{The conductivities for $\alpha<0$ and $\beta<0$. A spike
also emerges in the rightmost figure. }}
\end{figure}

Next we consider the electrical conductivity for non-positive
coupling parameters. Following the algebra in previous section, it
is straightforward to obtain the numerical results of the
conductivity for different coupling parameters. For convenience we
choose those values in TABEL\ref{table1} for $\alpha$ and $\beta$
and show the numerical results in FIG.8-FIG.11 separately. From
these figures, we notice that the ratio $\omega_g/T_c$ is unstable
which is similar to the case presented in section III. One
difference is that this ratio may shift to the region less than
$8$ when the coupling parameters are negative enough. Moreover, we
notice that there exist extra spikes that appear inside the gap as
shown in FIG.10 and FIG.11. These spikes emerge only when coupling
terms are negative enough, and such a phenomenon has also been
discussed in other holographic superconductor models when the mass
term is close to the BF bound\cite{Horowitz add1,Horowitz
add2,Wu2}.

By now we have investigated the superconductivity from the side of
bulk gravity. In the end of this section we remark that perhaps it
is instructive to understand this phenomenon from the side of
conformal field theory. Thanks to the AdS/CFT dictionary, the
correspondence between the couplings in the bulk and the couplings
on the boundary is manifest in quasi-topological
gravity\cite{Myers2}. On the boundary, the dual CFT is
characterized by central charges, $c$ and $a$, and flux
parameters, $t2$ and $t4$. Explicitly, these parameters can be
related to the coupling constants in the bulk as follows
\begin{eqnarray}\label{aaa}
\delta&=&\frac{c-a}{c}=\frac{4f_{\infty}(\alpha-3\beta f
_{\infty})}{1-2\alpha f_{\infty}-3\beta f_{\infty}^{2}},
\end{eqnarray}
\begin{eqnarray}
t2&=&\frac{24f_{\infty}(\alpha-87f_{\infty}\beta)}{1-2\alpha
f_{\infty}-3\beta f_{\infty}^{2}}\\
t4&=&\frac{3780f_{\infty}^{2}\beta}{1-2\alpha f_{\infty}-3\beta
f_{\infty}^{2}}\label{aab}
\end{eqnarray}
\begin{figure} \center{
\includegraphics[scale=0.6]{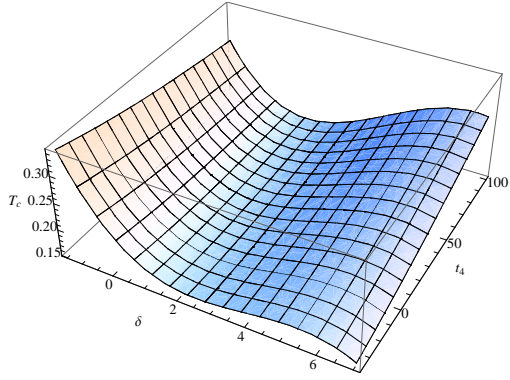}\hspace{0cm}
\caption{The critical temperature changes with $\delta$ and $t4$}}
\end{figure}
where $ f_{\infty}$ satisfies Eq.(\ref{c2}). Therefore, we may
establish a direct relation between the critical temperature of
the phase transition and the parameters in strongly coupled
system, namely $a$, $c$, $t_2$ and $t_4$. From
Eq.(\ref{aaa}-\ref{aab}), we know any change of the coupling
parameters in the gravity theory will lead to a change of
parameters $a$, $c$ and $t2$, $t4$ on the CFT side. However, since
there are only two free couplings in the gravity theory, two free
parameters is enough to describe the correlations on the CFT side.
For simplicity, we take $\delta$ and $t_4$ as the free parameters
and make a 3D plot to demonstrate how $T_c$ changes with the
running of $\delta$ and $t4$, which is illustrated in FIG. 12.
From this figure, we notice $T_c$ depends on $\delta$ more
sensitively, but in general we find that the dependence of $T_c$
on the parameters is not so simple as that on the gravity side.
Thus a clear understanding on this numerical result from the CFT
side is still missing. We expect further investigation would
disclose this with more details.

\section{Discussion and Conclusions}
In this paper we have constructed a (3+1) holographic
superconductor model in quasi-topological gravity which contains a
curvature-cubed interaction term. Firstly, we study the
condensation of the scalar field in the probe limit, and obtain
the relation between the condensation and the temperature. The
results obtained through an analytical approximation are
compatible with those through numerical analysis as well.
Secondly, we find higher curvature terms with both positive
coupling parameters suppress the phase transition such that the
 condensation becomes harder comparing with those in ordinary
Einstein's gravity theory. From a physical point of view, this is
because the positive correction terms in the quasi-topological
gravity which describe quantum fluctuations in the bulk will lead
to thermal fluctuations of the order parameter in the dual field
theory and thus suppress the occurrence of the spontaneous
breaking of the U(1) symmetry. Nevertheless, since we study the
holographic superconductor in a higher dimensional spacetime,
these quantum fluctuations are not strong enough to destroy the
condensation, in contrast to the (2+1) holographic superconductors
where the Coleman-Mermin-Wagner theorem is applicable\cite{CMW}.
Thirdly, we calculate the conductivity of holographic
superconductors numerically and find a running ratio
$\omega_g/T_c$ which becomes larger with the increase of couplings
parameters. Moreover, we find the cubic interaction term
suppresses the imaginary part. Especially, when the cubic coupling
$\beta$ is large enough, the line of imaginary part of the
conductivity is always below the one of real part. Finally, we
extend our discussion to the general case where the coupling
parameters may be non-positive. A general rule we found is that no
matter couplings are positive or negative, the critical
temperature always decreases when the couplings increase. However,
on the side of CFT it seems hard to figure out any general rule
which can describe its dependence on central charges and flux
parameters on the boundary. In addition, the ratio $\omega_g/T_c$
is always unstable but increasing when the coupling parameters
become larger.

Through this paper we have only considered the simplest s-wave
superconductors. Recent progress have shown that there also exist
holographic p-wave and d-wave superconductors\cite{p,d}. In these
models the order parameters are no longer scalar and the phase
transition corresponds to the symmetry breaking due to a
condensation of non-Abelian gauge fields. Some non-conventional
features which are different from s-wave ones have already been
disclosed. Our investigation on p-wave superconductor in
quasi-topological gravity is under progress and will be presented
elsewhere\cite{kll2}.

\begin{acknowledgments}
We are grateful to Li-Qing Fang, Gaston Giribet, Qiyuan Pan, Bin
Wang, Jian-Pin Wu, Hai-Qing Zhang, Hongbao Zhang and Chang-Chun
Zhong for reply and useful discussions. This work is partly
supported by NSFC(Nos.10663001,10875057), JiangXi SF(Nos. 0612036,
0612038), Fok Ying Tung Education Foundation(No. 111008), the key
project of Chinese Ministry of Education(No.208072) and Jiangxi
young scientists(JingGang Star) program. We also acknowledge the
support by the Program for Innovative Research Team of Nanchang
University.
\end{acknowledgments}


\begin{thebibliography}{99}
\bibitem{ADS}
J. M. Maldacena, The large N limit of superconformal field theories
and supergravity, Adv. Theor. Math. Phys. 2, 231 (1998).

\bibitem{WittenADS}
E. Witten, Anti-de Sitter space and holography, Adv. Theor. Math.
Phys. 2 (1998) 253¨C291, hep-th/9802150.

\bibitem{MaldacenaReviewADS}
O. Aharony, S. S. Gubser, J. M. Maldacena, H. Ooguri, and Y. Oz,
Large N field theories, string theory and gravity, Phys. Rept. 323
(2000) 183¨C386, hep-th/9905111.


\bibitem{Hartnoll1}
S. A. Hartnoll, P. K. Kovtun, M. Muller and S. Sachdev, Theory of
the Nernst effect near quantum phase transitions in condensed
matter, and in dyonic black holes, Phys. Rev. B 76, 144502 (2007),
arXiv:0706.3215 [cond-mat.str-el].

\bibitem{Hartnoll2}
S. A. Hartnoll and C. P. Herzog, Ohm¡¯s Law at strong coupling: S
duality and the cyclotron resonance, Phys. Rev. D 76, 106012 (2007),
arXiv:0706.3228 [hep-th].

\bibitem{Minic}
D. Minic and J. J. Heremans, High Temperature Superconductivity and
Effective Gravity, arXiv:0804.2880 [hep-th].

\bibitem{Hartnoll3}
S. A. Hartnoll, Lectures on holographic methods for condensed matter
physics, Class. Quant. Grav. 26, 224002 (2009) [arXiv:0903.3246].

\bibitem{Hartnoll4}
S. A. Hartnoll, Quantum Critical Dynamics from Black Holes,
arXiv:0909.3553.

\bibitem{McGreevy}
J. McGreevy, Holographic duality with a view toward many-body
physics, arXiv:0909.0518.

\bibitem{Gubser1}
S. S. Gubser, Phase transitions near black hole horizons, Class.
Quant. Grav. 22, 5121 (2005) ,arXiv:hep-th/0505189.
\bibitem{Gubser2}
S. S. Gubser, Breaking an Abelian gauge symmetry near a black hole
horizon, Phys. Rev. D 78, 065034 (2008) ,arXiv:0801.2977.

\bibitem{Hartnoll5}
S. A. Hartnoll, C. P. Herzog and G. T. Horowitz, Building an AdS/CFT
superconductor, Phys. Rev. Lett. 101:031601, 2008, arXiv:0803.3295.


\bibitem{Albash}
T. Albash and C. V. Johnson, A Holographic Superconductor in an
External Magnetic Field, JHEP0809:121,2008, arXiv:0804.3466.

\bibitem{Gubser3}
S. S. Gubser and S. S. Pufu, The gravity dual of a p-wave
superconductor, JHEP 0811:033,2008, arXiv:0805.2960.

\bibitem{Wen}
W. Y. Wen, Inhomogeneous magnetic field in AdS/CFT superconductor,
arXiv:0805.1550.

\bibitem{Gauntlett}
J. P. Gauntlett, J. Sonner and T. Wiseman, Holographic
superconductivity in M-Theory, Phys.Rev.Lett.103:151601,2009,
arXiv:0907.3796v3 [hep-th].

\bibitem{Gregory}
R. Gregory, S. Kanno, and J. Soda,Holographic Superconductors with
Higher Curvature Corrections, J. High Energy Phys. 0910, 010 (2009),
arXiv:0907.3203v3 [hep-th].

\bibitem{Sin}
S. J. Sin, S. S.  Xu and Y. Zhou, Holographic Superconductor for a
Lifshitz fixed point,  arXiv:0909.4857.

\bibitem{Cai}
R. G. Cai and H. Q. Zhang,  Holographic Superconductors with
Ho\v{r}ava-Lifshitz Black Holes, Phys.Rev.D81:066003,2010,
arXiv:0911.4867v2 [hep-th].

\bibitem{Ge}
X. H. Ge, B. Wang, S. F. Wu and G. H. Yang, Analytical study on
holographic superconductors in external magnetic field,
arXiv:1002.4901.

\bibitem{Pan}
 Q. Pan, B. Wang, E. Papantonopoulos, J. d. Oliveira and A. B. Pavan,
 Holographic Superconductors with various condensates in Einstein-Gauss-Bonnet gravity
 , Phys.Rev.D81:106007,2010,  arXiv:0912.2475;
 Q. Pan and B. Wang, General holographic superconductor models with
Gauss-Bonnet corrections, arXiv:1005.4743v1 [hep-th].

\bibitem{Wu}
J. P. Wu, The St\"{u}ckelberg Holographic Superconductors in
Constant External Magnetic Field,  arXiv:1006.0456.
\bibitem{Cai2}
 R. G. Cai, Z. Y. Nie and H. Q. Zhang, Holographic p-wave superconductors from Gauss-Bonnet
 gravity, arXiv:1007.3321v1 [hep-th].

\bibitem{Myers} R. C. Myers and B. Robinson, Black Holes in Quasi-topological
Gravity, arxiv:1003.5357v2.

\bibitem{oliva}J. Oliva and S. Ray, ¡°A new cubic theory of gravity in five dimensions: Black hole, Birkhoff¡¯s
theorem and C-function,¡± arXiv:1003.4773 [gr-qc].

\bibitem{ray}J. Oliva and S. Ray, ¡°A Classification of Six Derivative Lagrangians of Gravity and Static
Spherically Symmetric Solutions,¡± arXiv:1004.0737 [gr-qc].

\bibitem{sinha1}A. Sinha, ¡°On the new massive gravity and AdS/CFT,¡± JHEP 1006061 (2010), arXiv:1003.0683 [hep-th].

\bibitem{sinha2}A. Sinha, ¡° On higher derivative gravity, c-theorems and
cosmology,¡± arXiv:1008.4315[hep-th].

\bibitem{amsel}A. J. Amsel, D. Gorbonos, ¡°The Weak Gravity Conjecture and the Viscosity Bound with Six-Derivative
Corrections,¡± arXiv:1005.4718[hep-th].

\bibitem{Myers2}  R.C. Myers, M. F. Paulos and A. Sinha, Holographic studies of quasi-topological
gravity, arXiv:1004.2055v2 [hep-th].

\bibitem{Breitenlohner}
P. Breitenlohner and D. Z. Freedman, Positive Energy in anti-De
Sitter Backgrounds and Gauged Extended Supergravity, Phys. Lett.
B115 (1982) 197.
\bibitem{Son}
D. T. Son and A. O. Starinets, Minkowski-space correlators in
AdS/CFT correspondence: Recipe and applications, JHEP 0209, 042
(2002), arXiv:hep-th/0205051.
\bibitem{HorRob}
G. T. Horowitz and M. M. Roberts, Phys. Rev. D 78, 126008 (2008),
arXiv:0810.1077 [hep-th].
\bibitem{Horowitz add1}
G. T. Horowitz, Introduction to Holographic Superconductors, arXiv:1002.1722.
\bibitem{Horowitz add2}
G. T. Horowitz and M. M. Roberts, Holographic Superconductors with Various Condensates,
Phys. Rev. D 78 (2008) 126008 [arXiv:0810.1077].
\bibitem{Wu2}
J. P. Wu, Y. Cao, X. M. Kuang and W. J. Li, The 3+1 holographic superconductor with Weyl corrections
, arXiv:1010.1929v1 [hep-th].

\bibitem{Liu}
M. Brigante, H. Liu, R. C. Myers, S. Shenker and S. Yaida, Phys.Rev.D77:126006,2008, arXiv:0712.0805v3 [hep-th].

\bibitem{CMW}
D. Anninos, S. A. Hartnoll and N. Iqbal, Holography and the
Coleman-Mermin-Wagner theorem, arXiv:1005.1973v1 [hep-th].
\bibitem{p}
Gubser and S. S. Pufu, The gravity dual of a p-wave superconductor,
JHEP 0811, 033 (2008) [arXiv:0805.2960 [hep-th]].
\bibitem{d}
F. Benini, C. P. Herzog, R. Rahman and A. Yarom, Gauge gravity
duality for d-wave superconductors: prospects and challenges,
arXiv:1007.1981v1 [hep-th].

\bibitem{kll2} X.Kuang, W. Li and Y. Ling, Holographic p-wave
superconductor in quasi-topological gravity. In preparation.
\end{thebibliography}
\end{document}